\documentclass[10pt,conference]{IEEEtran}
\usepackage{cite}
\usepackage{amsmath,amssymb,amsfonts}
\usepackage{graphicx}
\usepackage{textcomp}
\usepackage[dvipsnames]{xcolor}
\usepackage[hyphens]{url}
\usepackage{fancyhdr}
\usepackage{hyperref}

\usepackage{multirow}
\usepackage{mathtools}
\usepackage{algorithm}
\usepackage{algpseudocode}
\usepackage{stfloats}
\usepackage{enumitem}
\usepackage{comment}
\usepackage{bbm}
\usepackage{nicefrac}
\usepackage{soul}

\usepackage{tikz} % circle numbers
\newcommand*\circled[1]{\tikz[baseline=(char.base)]{
            \node[shape=circle,draw,inner sep=1pt] (char) {#1};}}

\newcommand{\defcommenter}[2]{
  \expandafter\newcommand\csname #1\endcsname[1]{
  {\color{#2}[#1: ##1]}
  }
}
\makeatletter
\let\MYcaption\@makecaption
\makeatother

\usepackage[font=footnotesize]{subcaption}

\makeatletter
\let\@makecaption\MYcaption
\makeatother

\usepackage{setspace}
\title{Accelerating DNA Read Mapping with Digital Processing-in-Memory}
\author{Rotem~Ben-Hur,
       Orian~Leitersdorf,
       Ronny~Ronen,~\IEEEmembership{Fellow,~IEEE},  
       Lidor~Goldshmidt,
       Idan~Magram,
       Lior~Kaplum,
     \\  Leonid~Yavitz,
      and~Shahar~Kvatinsky,~\IEEEmembership{Member,~IEEE}% <-this % stops a space
}

\begin{document}
\maketitle

\thispagestyle{plain}
\pagestyle{plain}

%%%%%%%%%%%%%%%%%%%%%%%%%%%%%%%%%%%%%%%%
%%%%%%%% -- PAPER CONTENT STARTS -- %%%%%%%%%

\begin{abstract}

  Genome analysis has revolutionized fields such as personalized medicine and forensics. Modern sequencing machines generate vast amounts of fragmented strings of genome data called \textit{reads}. The alignment of these reads into a complete DNA sequence of an organism (the \textit{read mapping} process) requires extensive data transfer between processing units and memory, leading to execution bottlenecks. Prior studies have primarily focused on accelerating specific stages of the read-mapping task. Conversely, this paper introduces a holistic framework called DART-PIM that accelerates the entire read-mapping process. DART-PIM facilitates digital processing-in-memory (PIM) for an end-to-end acceleration of the entire read-mapping process, from indexing using a unique data organization schema to filtering and read alignment with an optimized Wagner–Fischer algorithm. A comprehensive performance evaluation with real genomic data shows that DART-PIM achieves a $\boldsymbol{5.7\times}$ and $\boldsymbol{257\times}$ improvement in throughput and a  $\boldsymbol{92\times}$ and $\boldsymbol{27\times}$ energy efficiency enhancement compared to state-of-the-art GPU and PIM implementations, respectively.

\end{abstract}

\section{Introduction}
\label{sec:introduction}

\par Genome analysis serves as the cornerstone of numerous scientific and medical breakthroughs, playing a vital role in personalized medicine~\cite{ginsburg2009genomic,chin2011cancer,ashley2016towards,hassan2022innovations} and advanced forensics~\cite{yang2014application,borsting2015next}. 
Genome sequencing machines~\cite{Illumina_NextSeq,Illumina_NovaSeq6000,Illumina_HiSeqX,Nanopore} produce vast volumes of short random fragmented strings of bases (A, C, G, T in DNA), commonly referred to as \textit{reads}, which are extracted from the longer original DNA~\cite{Onur2020review}. The process of reconstructing the original DNA sequence begins by approximating the location of each read in the overall genome and then using the reads at each location to deduce the most likely reference base.
This procedure involves complex algorithms with high memory requirements that constitute a bottleneck to the entire genome analysis process.

\par
\begin{figure}[!t]
    \centering
    \includegraphics[trim={0cm 0.1cm 0.1cm 0.1cm,width=1\linewidth,clip}, height=3.5cm,clip]{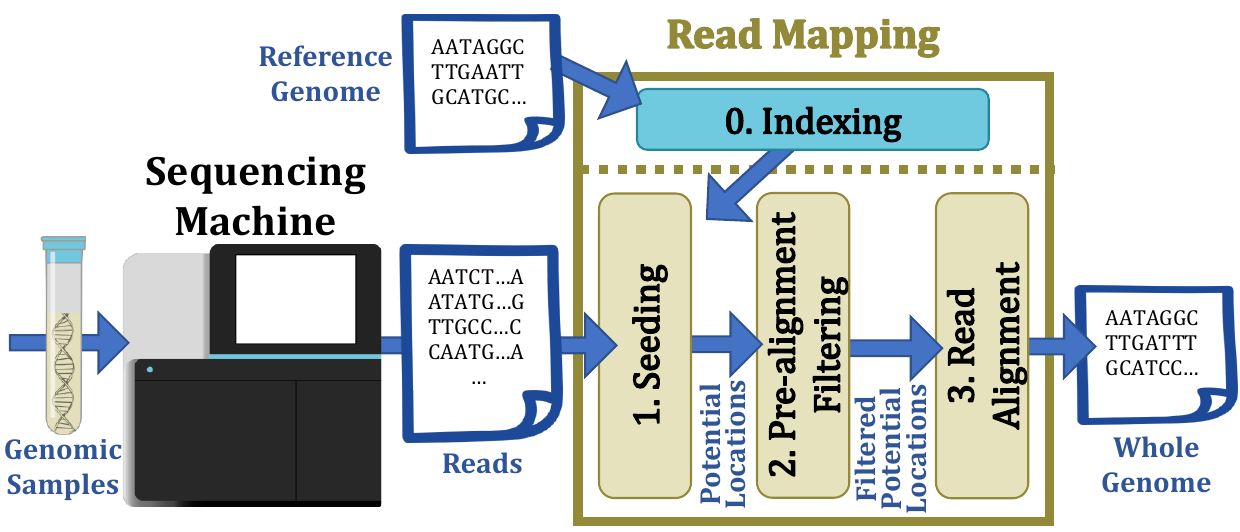}
    \caption{Genome sequence alignment process. The genomic samples are input into a \textbf{sequencing machine} that fragments them into small segments. The sequencing machine generates short strings known as \textbf{reads}, which are then processed during the \textbf{read-mapping} procedure. Read mapping involves an offline indexing stage followed by the online seeding, pre-alignment filtering, and read alignment stages.}
    \label{fig:Genome_analysis}
\end{figure}

 \textit{Sequence alignment} is a popular approach that localizes read fragments based on their similarity to a corresponding \textit{reference genome} of the target species.
This approach is feasible thanks to the high degree (above $99\%$) of resemblance between genomes of different specimens within the same species.
This computational process, called \textit{read mapping}~\cite{Onur2020review,canzar2015read_mapping}, involves offline \textit{indexing} of the reference genome (utilizing short representative fragments known as minimizers~\cite{minimizers}), followed by three consecutive online steps (illustrated in Figure~\ref{fig:Genome_analysis}): (1) \textit{Seeding} of potential locations (\textit{PLs}) in the reference genome based on read and reference minimizer similarity; (2) \textit{pre-alignment filtering} to eliminate false PLs; and (3) \textit{read alignment} that determines the most probable PL according to a similarity metric.
This latter stage is the most computationally intensive task, invoking string matching algorithms~\cite{hall1980ASM, ukkonen1985ASM, navarro2001ASM} for each read and PL within the complete genome.

\par State-of-the-art read alignment techniques predominantly employ a dynamic programming (\textit{DP})-based approach, such as the Needleman-Wunsch or Smith-Waterman algorithms~~\cite{levenshtein1966binary, needlemanWunsch1970, smithWaterman1981}. The complexity of these algorithms follows quadratic scaling in execution time and memory usage, thereby generating a significant latency and energy bottleneck~\cite{shouji, Onur2020review, huangfu2019medal, GenCache, darwin2018}. This read-mapping bottleneck has been exacerbated by the recent enhancements of read generation rates on the one hand and insufficient corresponding growth in computational power on the other~\cite{Onur2020review,hayden2014technology}.

\par To bridge the gap between read volumes and compute capabilities, prior work focused mainly on accelerating each read-mapping stage separately. This effort often prioritizes optimizing read alignment due to its significant computational demands, which can consume over 70\% of the read-mapping execution time~\cite{GenASM,RAPID_Gupta,GenCache,diab2023framework,Bioseal,FiltPIM,SeGraM}.
Read mapping, however, involves not only the processing of huge datasets, but also the frequent movements of huge volumes of data between the processor(s) and memory units, which increase both execution time and energy consumption~\cite{mutlu2019processing,Onur2020review}.
Since the data being transferred throughout the read-mapping process is roughly $100\times$ larger than the original input reads, the acceleration of computational tasks across the process is still bottlenecked by the data transfer between the processes.

\par Numerous studies have proposed alternative approaches to efficiently process large amounts of reads~\cite{GenASM, RAPID_Gupta, GenCache, diab2023framework,Bioseal,FiltPIM,SeGraM}. For instance, the state-of-the-art read mapper SeGraM~\cite{SeGraM} leverages near-memory computing~\cite{JohnReuben2017} within a 3D-stacked memory architecture to mitigate costly chip-to-chip data transfers. Consequently,  SeGraM's read alignment process alone achieves a $144\times$ speedup over existing CPU-executed software read mappers, alongside a $4.4\times$ power consumption reduction. Nevertheless, when handling the entire read-mapping process, SeGraM's performance demonstrates only a $2.5\times$ speedup for short reads over the existing CPU-based read mappers.

\par 
A promising solution to overcoming the data transfer bottleneck involves conducting read-mapping computations directly within the memory. This approach utilizes digital processing-in-memory (\textit{PIM}) techniques to enable high parallelism, eliminating the need for frequent data transfer between the memory and the processor~\cite{MUTLU201928}.
This paper presents \textit{DART-PIM}, a DNA read mApping acceleRaTor using Processing-In-Memory, which is a comprehensive framework for accelerating the entire read-mapping process. The uniqueness of DART-PIM lies in its ability to execute all read-mapping stages for a given read and PL inside a single memory crossbar array. The majority of computations are performed using the memory cells, thus circumventing the data transfer bottleneck. A small number of complementary operations are executed in adjacent RISC-V cores.
A key element for efficient computation in DART-PIM is careful organization of the reads and the reference genome within memory instances to minimize data transfer along all steps of read mapping.
Within these instances, the data allocation is optimized for maximal utilization of the inherent massive parallelism offered by PIM. 
Consequently, DART-PIM greatly enhances the performance and energy efficiency of the entire end-to-end read-mapping process. This paper makes the following main contributions:
\begin{enumerate}
    
    \item We propose a novel end-to-end accelerator architecture for the entire read-mapping process.
    
    \item We present a data organization technique for accelerated indexing and seeding of a reference genome.
    
    \item We develop a high-performance, memory efficient, in-memory pre-alignment filtering mechanism based on the linear Wagner-Fischer algorithm~\cite{WF}.
    
    \item We improved read alignment in-memory performance by enhancing the Wagner–Fischer algorithm with an affine-gap penalty and traceback capability.
    
    \item We build system-level simulators that use real genome datasets to ensure precise evaluation of DART-PIM's accuracy and performance, and compare it to state-of-the-art implementations (NVIDIA Parbricks~\cite{parabricks} and SeGraM~\cite{SeGraM}), demonstrating faster execution time ($5.7\times$ and $257\times$ improvement, respectively) and improved energy efficiency ($92\times$ and $27\times$, respectively).
    %\item Our suggested technique can be applied to other dynamic programming algorithms.
\end{enumerate}

%%%%%%%%%%%%%%%%%%%%%%%%%%%%%%%%%%%%%%%%%%%%%%%%%%%%%%%%%%%%%%%%%%%%%%%%%%%%%%%%%%%%%%%%%%%%%%%%%%%%%%%%%%%%%%%%%%%%%%%%%%%%%%%%%%
% BACKGROUND: Read Mapping and Its Limitations
%%%%%%%%%%%%%%%%%%%%%%%%%%%%%%%%%%%%%%%%%%%%%%%%%%%%%%%%%%%%%%%%%%%%%%%%%%%%%%%%%%%%%%%%%%%%%%%%%%%%%%%%%%%%%%%%%%%%%%%%%%%%%%%%%%

\section{Background: Read Mapping and Its Limitations}
\label{sec:seq_align}

The read-mapping process comprises three sequential online stages, preceded by an offline indexing procedure. The classical indexing approach generates a database that matches any sub-string of length $k$ (typically, $12$) appearing in the reference genome (called \textit{k-mer}) with pointers to all locations of this sub-string within the genome~\cite{k-mers}. Although lossless, the practicality of this approach is limited due to its extensive memory demands. A more practical approach indexes the genome to select \textit{minimizers}~\cite{minimizers}, rather than k-mers, which reduces the number of k-mers associated with each reference segment while achieving accurate compressed representation.

Once the indexing stage is completed, the database is stored and each new DNA sample can be sequenced by employing the following three steps: (1) seeding, (2) pre-alignment filtering, and (3) read alignment. These steps are performed sequentially for any new read (typically, strings comprising $100$ to $2000$ bases) entering the system. 
The purpose of the seeding stage is to identify PLs of the read within the reference genome. The read is processed in sliding windows of length $W+k-1$, where $W$ is the window size (typically $W=30$), and each window is represented by its minimizer. The PLs of the read are defined as the set of addresses of all its (unique) minimizers in the reference genome (taken from the indexing database). 

Since the sets of PLs are typically large, a filtering procedure is required to reduce the number of PLs per read. Pre-alignment filtering commonly employs heuristic techniques to discard PLs based on dissimilarity between the read and a reference genome segment. For instance, a popular filtering approach, called the base count filter~\cite{BaseCountFilter}, compares histograms of bases for a read and a corresponding segment (eliminating $68\%$ of PLs on average).

The main step of read mapping involves optimal sequence alignment algorithms for read alignment. Since the goal is to determine the most accurate location of each read within the genome, it commonly involves computationally-intensive dynamic-programming algorithms such as the Smith-Waterman (\textit{SW}) algorithm~\cite{SW} or the Needleman-Wunsch (\textit{NW}) algorithm~\cite{NW}.
While optimal sequence alignment is the obvious candidate for hardware acceleration due to its compute-bound nature, the primary portion of energy and execution time stems from data transfer between the memory and computing cores throughout the sequencing process. 

Based on in-house simulations we conducted with a human reference genome ($0.8$GB in size), seeding generates an average of $1000$ ($32$ bits) PLs per read, which amounts to $1556$GB for a typical amount of $389$M reads of length $150$ ($14.6$GB in size). Note that while the input data volume to the seeding stage is $14.6$GB, it generates approximately a $100$ times larger output data volume, which has to be transferred to the memory and back to a pre-alignment filtering accelerator. This data transfer is the system's primary bottleneck, even with significant acceleration of each individual step. Motivated by this observation, this paper proposes a solution that executes all stages of read mapping within the same memory crossbar array, thereby eliminating the need for data transfers between the different stages.

%%%%%%%%%%%%%%%%%%%%%%%%%%%%%%%%%%%%%%%%%%%%%%%%%%%%%%%%%%%%%%%%%%%%%%%%%%%%%%%%%%%%%%%%%%%%%%%%%%%%%%%%%%%%%%%%%%%%%%%%%%%%%%%%%%
% Read Mapping Algorithms 
%%%%%%%%%%%%%%%%%%%%%%%%%%%%%%%%%%%%%%%%%%%%%%%%%%%%%%%%%%%%%%%%%%%%%%%%%%%%%%%%%%%%%%%%%%%%%%%%%%%%%%%%%%%%%%%%%%%%%%%%%%%%%%%%%%

\section{Modified Read-Mapping Algorithms Used in DART-PIM}
\label{sec:WF_with_PIM}

This section describes the read-mapping algorithms that are employed in DART-PIM for efficient processing in-memory. 
The basic algorithms are the \textit{Linear Wagner-Fischer} and \textit{Affine Wagner-Fischer} algorithms. The former is used for pre-alignment filtering and the latter for read alignment.
The preceding stages of read mapping, i.e., indexing and seeding, only require that the data be carefully arranged within the memory prior to pre-alignment filtering.

State-of-the-art read mappers commonly employ either the Smith-Waterman (\textit{SW}) algorithm~\cite{SW} or the Needleman-Wunsch (\textit{NW}) algorithm~\cite{NW}. Since the compared strings in DNA sequencing setting are typically similar, these algorithms, which consider sub-string matches, require relatively large bit-width representation (e.g., $8$ bits) for storing the similarity score. Hence, supporting read mapping by processing it within the memory may impose a relatively large memory footprint or restrict the sequence length that can be processed efficiently.

To enhance read mapping efficiency, we propose an alternative algorithmic approach that adopts the Wagner-Fischer (WF) algorithm~\cite{WF}, originally used for edit distance computation. We modify the algorithm to output not only the distance metric, but also the traceback (i.e., the aligned sequence itself). Since the WF algorithm counts mismatches, rather than matches, it requires fewer bits (e.g., $3$) per distance, thereby being a better fit for processing-in-memory.

We use two variants of the WF algorithm. The first variant, used for pre-alignment filtering, is an adaptation of the classical WF algorithm.
We call this the \textit{Linear} WF algorithm. 
The second variant, used for read alignment, extends the linear WF by enhancing the WF algorithm by (1) accounting for the affine-gap penalty of the SW and NW algorithms, and (2) enabling traceback to reconstruct the aligned sequence. We refer to this as the \textit{Affine} WF algorithm. The affine-gap penalty combines constant and linear gap penalties and is commonly used in biological applications~\cite{AffineGapPenalty1,AffineGapPenalty2}.

\subsection{Linear Wagner-Fischer Algorithm}
Given two input strings $S_1$ and $S_2$ (with respective lengths $n$ and $m$), the linear WF algorithm constructs the \textit{WF distance matrix}, $D$, according to the conditions of the WF algorithm, with a size of $(n+1) \times (m+1)$:
\begin{enumerate}
    \item Initialize the first row and first column as follows:
    {\footnotesize
    \begin{equation}
    \begin{aligned}
        D_{i,0} & = \sum_{k=1}^{i} w_{del}, \text{ for } 1 \leqslant i \leqslant n, \\
        D_{0,j} & = \sum_{k=1}^{j} w_{ins}, \text{ for } 1\leqslant j \leqslant m.
    \end{aligned}
    \end{equation}}
    \item Fill the matrix $D$: If $S_1(i)=S_2(j)$, then $D_{i,j} = D_{i-1,j-1}$; otherwise,
    {\footnotesize
    \begin{equation}\label{eq:linearWF_D}
        D_{i,j} = \min
                \begin{rcases}\begin{dcases}
                    D_{i-1,j}+w_{del}, \\
                    \ D_{i,j-1}+w_{ins}, \\
                    \ \ D_{i-1,j-1}+w_{sub}
                \end{dcases}\end{rcases}.
    \end{equation}}
\end{enumerate}
    where $w_{del}$, $w_{ins}$, and $w_{sub}$ are the costs of a deletion, an insertion, and a substitution, respectively. The cost values, as well as additional WF parameters, appear in Table~\ref{tab:dart-pim-params}.

\subsection{Modified Affine Wagner-Fischer Algorithm}

The affine WF algorithm enhances mapping accuracy (at the cost of complexity) by using a penalty that is represented as $w_{op}+w_{ex}(L-1)$, where $w_{op}$ denotes the cost of opening a gap, $w_{ex}$ represents the cost of extending an existing gap while $L$ corresponds to the gap length.
Since this distance measure achieves higher accuracy at the cost of higher complexity (larger execution time and memory storage), we perform the linear WF algorithm for each PL for pre-alignment filtering, and employ the affine WF only for selected locations based on their linear WF distance. 
The explicit affine-gap WF matrix is given by

{\footnotesize
\begin{equation}\label{eq:affineWF_D}
    D_{i,j} = 
              \begin{cases}
                D_{i-1,j-1}, & S_1(i)=S_2(j)\\[10pt]
                \min
                    \begin{rcases}\begin{dcases}
                        M1_{i,j}, & \text{(ins)}\\
                        M2_{i,j}, & \text{(del)}\\ 
                        D_{i-1,j-1} + w_{sub} & \text{(sub)}
                    \end{dcases}\end{rcases}
                , & \text{otherwise,}
              \end{cases}
\end{equation}}
where 
{\footnotesize
\begin{equation}\label{eq:affineWF_M1}
    M1_{i,j} = 
                \min
                \begin{rcases}\begin{dcases}
                    M1_{i-1,j} + w_{ex}, & \text{(extend S1 gap)} \\                    
                    D_{i-1,j} + w_{op} + w_{ex} & \text{(open S1 gap)}
                \end{dcases}\end{rcases}
\end{equation}}
and
{\footnotesize
\begin{equation}\label{eq:affineWF_M2}
    M2_{i,j} = 
                \min
                \begin{rcases}\begin{dcases}
                    M2_{i,j-1} + w_{ex}, & \text{(extend S2 gap)}\\
                    D_{i,j-1} + w_{op} + w_{ex} & \text{(open S2 gap)}
                \end{dcases}\end{rcases}
\end{equation}}
Due to the iterative nature of the WF algorithms, the $(i,j)$-th matrix value $D_{i,j}$ originates from one of three possible preceding cells (see Eq.~\ref{eq:affineWF_D}), the $(i,j)$-th matrix values $M1_{i,j}$, $M2_{i,j}$ originate from one of two possible preceding cells (see Eq.~\ref{eq:affineWF_M1} and~\ref{eq:affineWF_M2}). The affine WF saves the ``directions'' of the predecessors ($4$-bit representation) for each matrix cell. The optimal sequence alignment can, therefore, be inferred without having to save the entire matrix for traceback calculation.

%%%%%%%%%%%%%%%%%%%%%%%%%%%%%%%%%%%%%%%%%%%%%%%%%%%%%%%%%%%%%%%%%%%%%%%%%%%%%%%%%%%%%%%%%%%%%%%%%%%%%%%%%%%%%%%%%%%%%%%%%%%%%%%%%%
% In-Memory Implementation of Read Mapping Algorithms %%%%%%%%%%%%%%%%%%%%%%%%%%%%%%%%%%%%%%%%%%%%%%%%%%%%%%%%%%%%%%%%%%%%%%%%%%%%%%%%%%%%%%%%%%%%%%%%%%%%%%%%%%%%%%%%%%%%%%%%%%%%%%%%%%

\section{In-Memory Execution of Read-Mapping Algorithms}
The read mapping execution in DART-PIM leverages digital processing-in-memory (PIM) technologies. Emerging digital PIM architectures (such as DRAM PIM~\cite{ComputeDRAM, SIMDRAM, DRISA, Ambit}, memristive PIM~\cite{IMPLY, FELIX, MAGIC, MemristiveLogic, Nishil_tnano, RACER2021} and SRAM PIM~\cite{NeuralCache, DualityCache}) enable massive parallel bitwise operations within memory by integrating logic circuits into the same physical devices used for memory. For explanation and evaluation, we chose to employ bulk bitwise PIM~\cite{BenHPCA,BenTETC}, using memristive memory technologies~\cite{RRAM,PCM,STT-MRAM}. However, other digital PIM technologies can also be used. This section outlines the basic principles of our approach using the MAGIC NOR gate~\cite{MAGIC}, and explains how a WF iteration is implemented within a single memory crossbar row.

\subsection{Memristive Processing-In-Memory}
\label{sec:PIM}

% Memristive devices are capable of altering their resistance when electrically stimulated. These devices are advantageous for constructing memories thanks to their small form factor and dense arrangement in crossbar arrays~\cite{HighDensityReRAM,HighDensityPCM}. Additionally, their nonvolatile nature eliminates the need for memory refresh operations, reducing idle power consumption~\cite{Xu2015crossbar}. 
% Nonvolatile memristive memory technologies include resistive RAM (RRAM)~\cite{RRAM}, phase change memory (PCM)~\cite{PCM, 3DXpoint}, and magnetic tunnel junctions (MTJs)~\cite{STT-MRAM}.

Memristive memories can be used as nonvolatile memories, where each cell within the memory crossbar array consists of a single memristive device whose resistance encodes its logical state.
When writing or reading data, appropriate voltages are applied across the desired wordlines and bitlines to modify/sense the resistance of the target memristive devices~\cite{Xu2015crossbar}.
These memories can perform computations using the memory cells themselves~\cite{Borghetti2010, MAGIC, Gupta_Felix, Barak_VCM, Natan_X-MAGIC}.
Although the reliability of memristive logic is still challenging, specific strategies have been developed~\cite{Orian_DAC, Orian_ICECS, wear_leveling} to substantially improve their correct execution and lifetime. 

\begin{figure}
    \centering
    \includegraphics[width=\linewidth,clip]{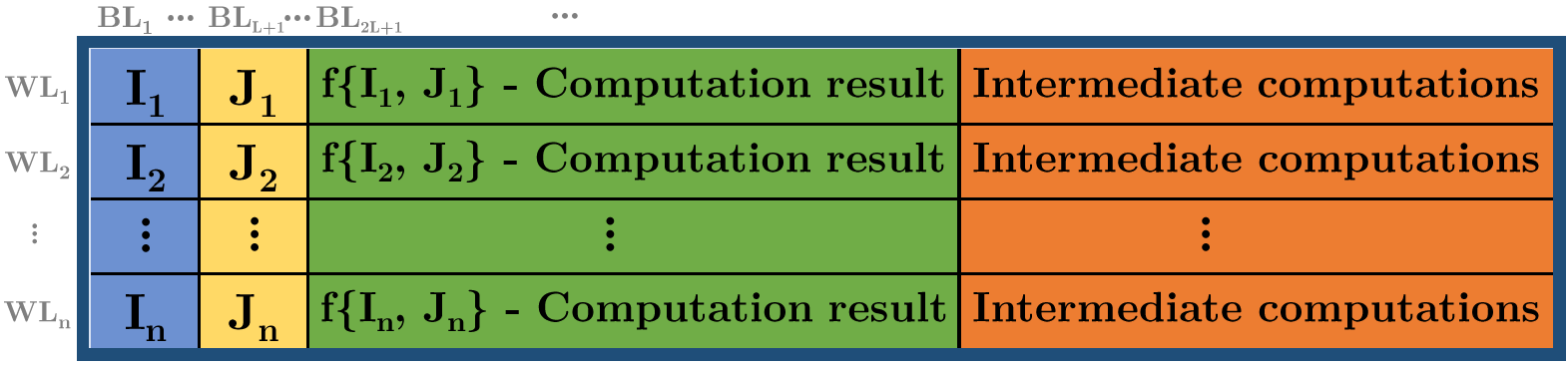}
    \caption{Area allocation within a memory crossbar array: each row conducts a series of logical operations, starting from $L$-bits inputs $I_x$ (blue) and $J_x$ (yellow), using intermediate results (orange), to obtain the final output stored in the memory (green). All $n$ rows perform the same logical operations (over different inputs) concurrently. $WL_x$ and $BL_x$ are, respectively, wordline (row) and bitline (column) number x.}
    \label{fig:MAGIC_sequence}
\end{figure}

One notable technique for conducting logic operations within an RRAM~\cite{RRAM,Xu2015crossbar,HighDensityReRAM} memristive array is the MAGIC NOR method~\cite{MAGIC}. 
This method involves applying voltages to two input memory cells and one output cell in the same row. The output memory cell changes its state according to the NOR function applied to the logical states of the two inputs.
The structure of MAGIC NOR allows parallel execution of multiple gates, each in a different row, as long as their columns are aligned. Tools such as~\cite{SIMPLER,SIMPLE} can determine the latency-optimal sequence of such operations to accomplish a desired computation.
This sequence is implemented using cells within the same row, operating sequentially. When all available cells become occupied by the outputs, the unused cells are initialized and reused for the remaining computations. An illustration of this procedure is depicted in Figure~\ref{fig:MAGIC_sequence}.

Implementing read mapping using PIM requires various logical operations. These operations and their corresponding optimized number of clock cycles are listed in Table~\ref{tab:MAGIC_based_opeartions}.
% The hardware implementation of these operations was optimized for MAGIC NOR in-memory compute.
%A more detailed discussion of these operations and their implementation is included in the paper's GitHub repository~\cite{DART-PIM_Github}.

\begin{table}[]
    \scriptsize
    \centering
    \caption{Summary of execution cycles for MAGIC-NOR-based operations with $N$-bit operands (unless specified otherwise).}
    \label{tab:MAGIC_based_opeartions}
    \begin{tabular}{||c | c||}
    \hline
        \textbf{Logical Operation} & \textbf{\# Cycles}  \\
    \hline \hline
        AND & $3N$ \\
    \hline
        XNOR & $4N$ \\
    \hline
        XOR & $5N$ \\
    \hline
        Copy & $1+N$ \\
    \hline
        Addition of two $N$-bit in-memory numbers & $9N$ \\ %Using a single-bit full adders
    \hline
        Addition of an $N$-bit and a single-bit in-memory numbers & $5N$ \\ %Using a single-bit half adders
    \hline
        Addition of one in-memory number and a constant & $5N$ \\
    \hline
        Subtraction of two in-memory numbers & $9N$ \\
    \hline
        Mux between two in-memory numbers & $3N+1$ \\
    \hline
        Minimum of two in-memory numbers & $12N+1$ \\
    \hline
    \end{tabular}
    
\end{table}

\subsection{Wagner-Fischer Algorithm in a Single Crossbar Row}
\label{subsec:single_row_WF}

Maximizing the parallelism of bulk-bitwise PIM is crucial in DART-PIM. Full potential is achieved when an entire WF algorithm is executed within a single crossbar row.
In this configuration, each row processes a separate instance of the WF algorithm, with different subsets of columns dedicated to specific algorithmic steps. This setup allows simultaneous operations on multiple reads/PLs across the crossbar's columns.

\begin{figure*}
    \centering
    \includegraphics[trim={0.1cm 0cm 0.1cm 0.1cm},width=1\linewidth,clip]{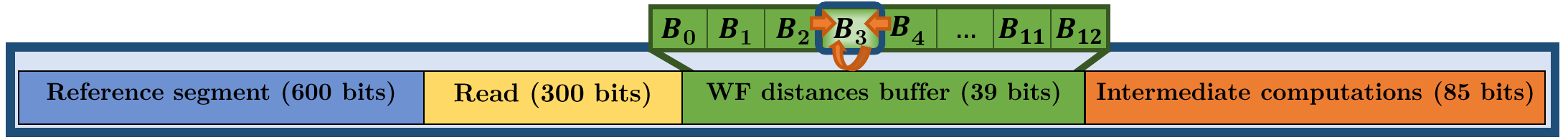}
    \caption{Mapping of a linear WF matrix calculation into a single crossbar row for $eth=6$. The reference segment (blue) and read (yellow) are the computation inputs. Only $2eth+1$ WF distances are needed at any point (green). To compute the current WF distance, only the distances in adjacent cells (storing the top and left WF matrix distances) and the previous value of the current cell (storing the top-left WF matrix distance) are required.
    The intermediate results generated while computing the distances are stored in temporary row cells (orange), and due to limited number of cells, are re-used when necessary.
    The total number of bits in the row is 1024.}
    \label{fig:single_row_WF}
\end{figure*}

\begin{figure}
    \centering
    \includegraphics[width=0.7\linewidth,clip]{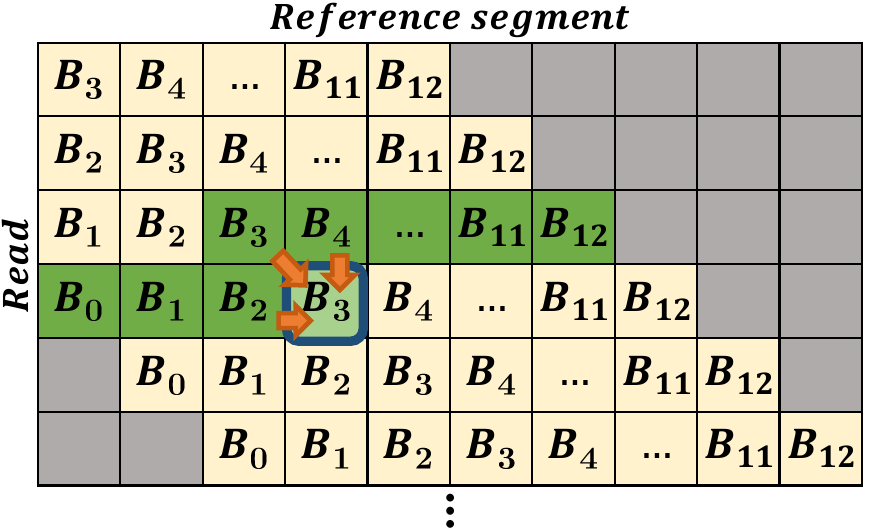}
    \caption{Example of the mapping of a banded WF matrix (with $eth=6$) onto the WF distances buffer in a single crossbar row (contained $2eth+1$ cells). The computed cell is denoted by light green ``$B_3$'' (location $(4,4)$). 
    The remaining green cells are stored within the WF distances buffer  during computation. The computation is performed using only top, top-left, and left cells (marked with orange arrows). Upon completion, the computed value replaces the target cell's value in the WF distances buffer. Note that gray cells are not computed as they represent distances larger than $eth$.}
    \label{fig:single_row_WF_matrix}
\end{figure}

To meet the single-row requirement, we truncate the maximal allowed WF matrix values to $eth$, beyond which the strings are considered too different and the value is saturated. The selection of $eth$ trades off alignment accuracy for complexity reduction, as it allows to compute in each row only $2eth+1$ unsaturated cells around the diagonal~\cite{banded_SW}. We assign $3$-bits per value for linear WF ($5$-bit for affine WF), setting $eth$ to $6$ ($31$ for affine WF). This maintains accuracy surpassing even the commonly used $eth=5$ for linear SW~\cite{Illumina_HiSeqX}.

Algorithm~\ref{alg:single_cell_WF} details the steps to calculate a single WF matrix cell using MAGIC NOR operations within a single memory row. Most row cells store the read ($S_1$) and reference segment ($S_2$), while other cells are used for the computation. WF distances have a designated cell subset (\textit{WF distances buffer}), with remaining cells for intermediate computations. This allocation is shown in Figure~\ref{fig:single_row_WF}, and WF matrix cell mapping in Figure~\ref{fig:single_row_WF_matrix}. This core process is optimized for a minimal number of clock cycles.

\begin{algorithm}
    \caption{$D_{i,j}$ calculation for the linear WF with MAGIC ({\color{CadetBlue} \# MAGIC operations (ops) for a $b$ bit cell, derived from Table~\ref{tab:MAGIC_based_opeartions}}).}
    \label{alg:single_cell_WF}
    {\footnotesize
    \begin{algorithmic}[1]
        \Statex \textbf{Input}: String characters $S_1(i), S_2(j)$ and previous WF values $D_{i-1,j}, D_{i,j-1}, D_{i-1,j-1}$
        \Statex \textbf{Output}: Next WF value $D_{i,j}$
        \State $X = \min\left\{ D_{i-1,j}, D_{i,j-1}\right\}$, $Y = \min(X,D_{i-1,j-1})$ 
        \Statex \Comment{{\color{CadetBlue} $2\cdot13b$ ops (adding two $b$-bit numbers)}}
        \State $Z=Y+1$ \Comment{$w_{del}=w_{ins}=w_{sub}=1$}
        \Statex \Comment{{\color{CadetBlue} $5b$ ops (adding $b$-bit and single bit numbers)}}
        \State Set $S1 = 7$ if Y=7, and $0$ otherwise
        \Statex \Comment{MUX1 select, {\color{CadetBlue} $6$ ops (two single-bit ANDs)}}    
        \State Set $MUX1=Y$ if $S1=1$, and $Z$ otherwise
        \Statex \Comment{{\color{CadetBlue} $3b+1$ ops}}
        \State Set $S2=1$ if $S_1(i)=S_2(j)$, and $0$ otherwise 
        \Statex \Comment{MUX2 select, {\color{CadetBlue} $11$ ops (two XNORs + single AND)}}
        \State \Return MUX2 value as $D_{i-1,j-1}$ if S2=1, and MUX1 otherwise
        \Comment{{\color{CadetBlue} $3b+1$ ops}}
        \Statex \Comment{{\color{CadetBlue} Total: $37b+19$ ops for a single cell}}
    \end{algorithmic}}
\end{algorithm}

\begin{algorithm}
    \caption{The linear WF matrix calculation with $2eth+1$ buffer size.}
    \label{alg:single_row_WF}
    {\footnotesize
    \begin{algorithmic}[1]
        \Statex \textbf{Input}: Strings $S_1$ (read of length $N$) and $S_2$ (reference of length $M$) \Comment{$eth$ parameter set to $6$}
        \Statex \textbf{Output}: Linear WF distance between $S_1$ and $S_2$
        \State Initialize to zero an array WFd of size $2eth+1$ and $3$-bit cells \Comment{WF distances buffer}
        \For {$i=0$ to $N-1$}
            \For {$j=0$ to $2eth$}
                \State Set $WFd[j]$ to one of the following:
                \If {($j==0$)} \Comment{left edge of WFd}
                    \begin{equation*}
                    \begin{aligned}
                        \min\left\{ 
                            \textnormal{WFd}[j] + \mathbbm{1}_{S_2[i] \neq S_1[i+j]}, 
                            \textnormal{WFd}[j + 1] + 1
                            \right\}
                    \end{aligned}
                    \end{equation*}
                \ElsIf{($j==2eth$)} \Comment{right edge of WFd}
                    \begin{equation*}
                        \min\left\{\textnormal{WFd}[j] + \mathbbm{1}_{S_2[i] \neq S_1[i+j]}, \textnormal{WFd}[j - 1] + 1\right\}
                    \end{equation*}
                \Else
                    \begin{equation*}
                        \min \begin{rcases}\begin{dcases}
                            \textnormal{WFd}[j] + \mathbbm{1}_{S_2[i] \neq S_1[i+j]}, 
                            \quad\textnormal{WFd}[j - 1] + 1, \\
                            \quad\quad\quad\quad\quad \textnormal{WFd}[j + 1] + 1 
                         \end{dcases}\end{rcases}
                    \end{equation*}
                \EndIf
                \State $\textnormal{WFd}[j] = \min\left\{\textnormal{WFd}[j], eth+1\right\}$
            \EndFor
        \EndFor
        \State \Return $\textnormal{WFd}[eth]$
    \end{algorithmic}}
\end{algorithm}

To fit all WF computations in a single row, we minimize crossbar row bits for input data (strings $S_1$ and $S_2$) and computations. Each WF matrix cell requires only its upper, upper-left, and left neighbors. 
The cells are computed sequentially within the crossbar row, where only the previous cells required for the present computation are stored at any given time, until the final result in the lower-left cell is reached. For linear WF, the aligned string is not needed, so traceback is not required.
The complete WF matrix computation using a buffer of $2eth+1$ cells is provided in Algorithm~\ref{alg:single_row_WF}.

Similarly, the affine WF computation involves three matrices, each with $2eth+1$ cells per row. 
To reconstruct the aligned string (traceback), we track the edits path by storing the direction of the origin cell for each matrix cell. 
According to Eq.~(\ref{eq:affineWF_D}), two bits are needed per cell for the four origin directions. Matrices M1 (Eq.~\ref{eq:affineWF_M1}) and M2 (Eq.~\ref{eq:affineWF_M2}), require one bit per cell for their two origin directions. Traceback information is stored in designated rows within the crossbar, occupying $7\times$ more rows than used for computation.

These WF optimizations reduce memory footprint and computational complexity by decreasing the bit-width per WF cell and considering fewer sub/super-diagonals.
This lowers the linear WF algorithm latency by $2.8\times$ compared to SW, enabling efficient computation of a full matrix within a single memristive crossbar row (versus two rows in SW).

%%%%%%%%%%%%%%%%%%%%%%%%%%%%%%%%%%%%%%%%%%%%%%%%%%%%%%%%%%%%%%%%%%%%%%%%%%%%%%%%%%%%%%%%%%%%%%%%%%%%%%%%%%%%%%%%%%%%%%%%%%%%%%%%%%
% In-memory DNA Sequencing 
%%%%%%%%%%%%%%%%%%%%%%%%%%%%%%%%%%%%%%%%%%%%%%%%%%%%%%%%%%%%%%%%%%%%%%%%%%%%%%%%%%%%%%%%%%%%%%%%%%%%%%%%%%%%%%%%%%%%%%%%%%%%%%%%%%
\section{DART-PIM: DNA Read-Mapping Accelerator}
\label{sec:DART-PIM_arch}

\par This section presents the DART-PIM architecture, integrating bulk-bitwise memristive PIM with RISC-V cores. We commence by describing DART-PIM's structure and properties, then explain each stage of read mapping using DART-PIM.

\subsection{DART-PIM Architecture}
\label{subsec:DART_arch}

\begin{figure}
    \centering
    \includegraphics[trim={0.1cm 0.1cm 0.1cm 0cm},width=1\linewidth,clip]{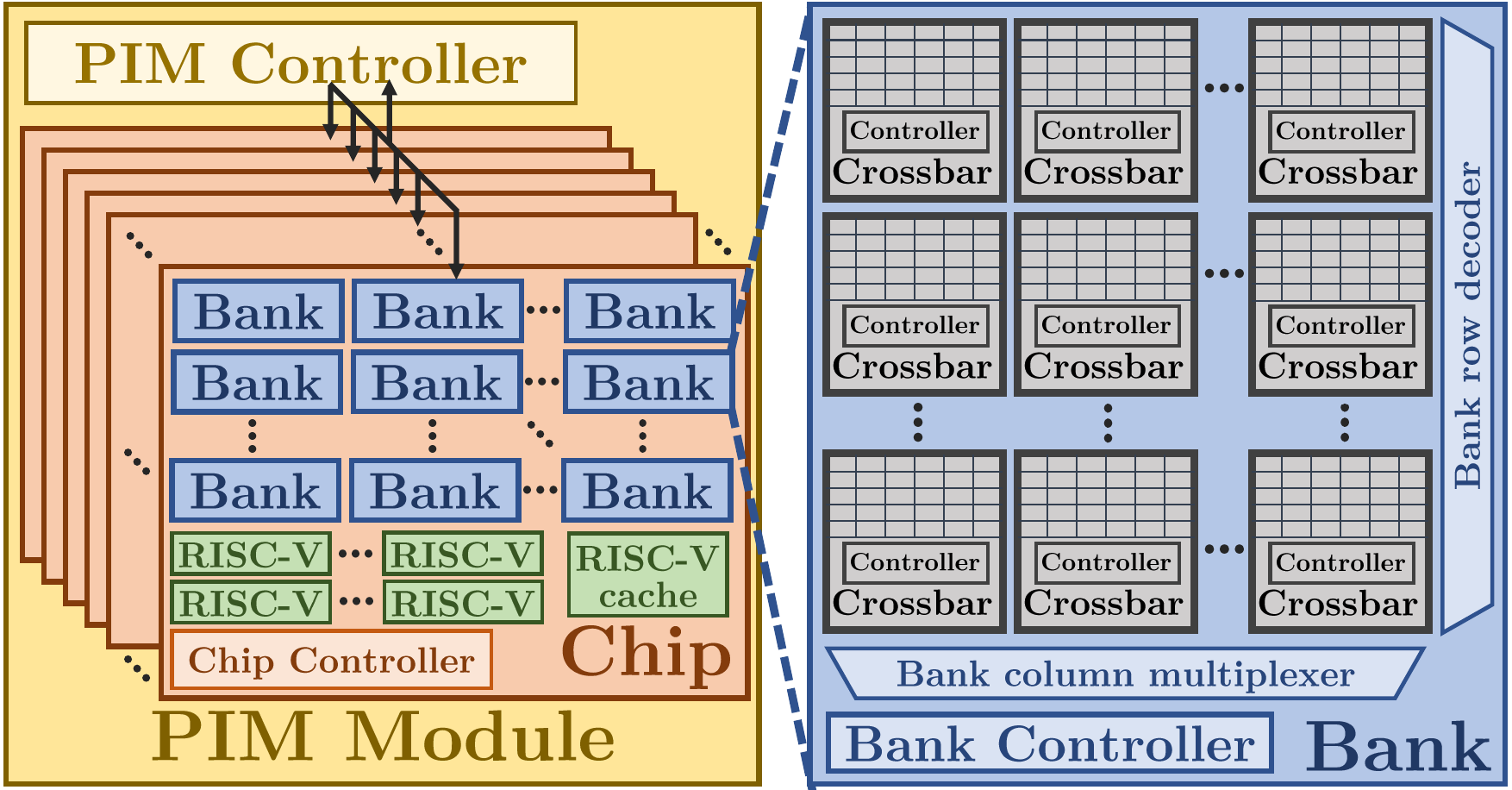}
    \caption{DART-PIM architecture featuring RISC-V cores and computing memristive memories. The memory consists of a single PIM module, which contains RISC-V cores with private L1 cache memories integrated within the memristive memory chips. Each chip is divided into banks, each equipped with a dedicated controller. Each crossbar in the bank is responsible for a single reference minimizer.}
    \label{fig:DART_arch}
\end{figure}

The DART-PIM architecture is based on a PIM module, which includes a PIM controller managing  multiple memristive chips, similar to standard DRAM main memories~\cite{jacob2010memory} (where a PIM module is analogous to a DRAM rank).
Each PIM module incorporates several memory chips with two primary components: (1) The DART-PIM memory (\textit{DP-memory}) -- A set of banks with a controller and multiple crossbars~\cite{jacob2010memory, Nishil_tnano}, and (2) the DART-PIM RISC-V (\textit{DP-RISC-V}) -- standard RISC-V processors performing a small portion ($0.16\%$) of the affine WF instances, with one processor (the \textit{main RISC-V}) managing data arrangement and high-level processing instructions for all DP-memory cores.
The hierarchical structure of DART-PIM is illustrated in Figure~\ref{fig:DART_arch} with specific architectural parameters given in Table~\ref{tab:configurations}.

\begin{table}[]
    \centering
    \caption{Summary of DART-PIM architecture configuration.}
    \label{tab:configurations}
    \scriptsize
    \begin{tabular}{|c|c|}
        \hline
        \textbf{Module property} & \textbf{Value} \\
        \hline
        Total memory capacity & $256$GB \\ 
        \hline
        $\#$ PIM Module & $1$ \\ 
        \hline
        $\#$ Chips per PIM module & $32$ \\ 
        \hline
        $\#$ Banks per chip & $512$ \\ 
        \hline
        $\#$ Crossbars per bank & $512$ \\ 
        \hline
        $\#$ Columns / rows / Bytes per crossbar  & $1024$ / $256$ / $32$KB \\ 
        \hline
        $\#$ RISC-V cores per chip & $4$ \\
        \hline
        Cache capacity per chip & $128$KB \\
        \hline
        RISC-V to memory banks BUS width & $512$ bits \\
        \hline
    \end{tabular}
\end{table}

In addition to standard read/write capabilities, the memristive crossbars in DART-PIM support PIM operations with dedicated per-chip controllers. Each chip controller oversees multiple banks and their crossbars. Upon initiating a PIM request, the information is transmitted to the target chip in a write operation format~\cite{Nishil_CONCEPT}. The designated chip controllers then issue the necessary sequence of MAGIC NOR operations to all associated banks and crossbars. Since all crossbars execute identical tasks, the controllers are quite simple, reducing system area and energy consumption.

\begin{figure}
    \centering
    \includegraphics[trim={0.1cm 0.1cm 0.1cm 0cm},width=\linewidth,clip]{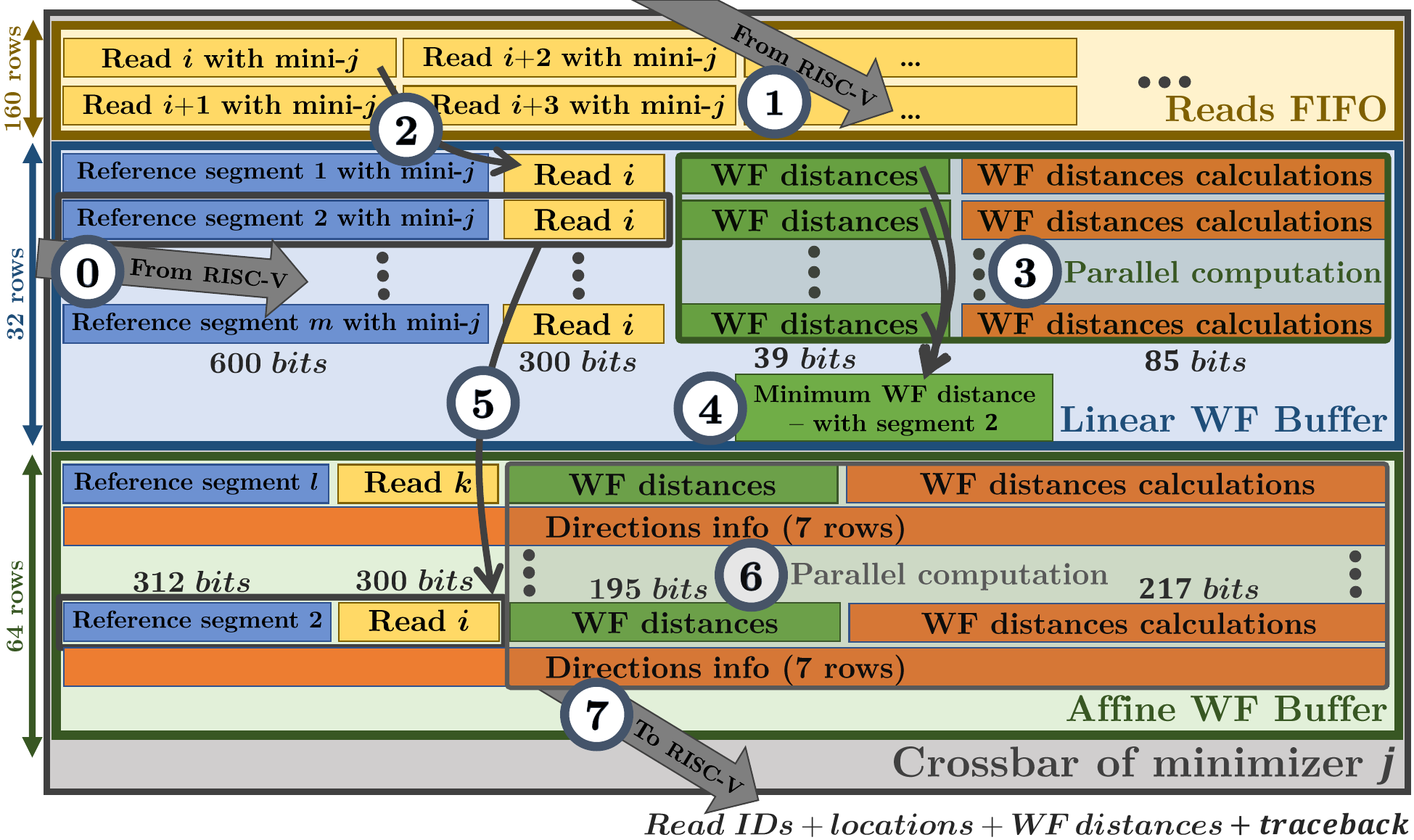}
    \caption{A memory crossbar with 256 rows and 1024 columns, used for the execution of the WF algorithm between reads and reference segments with minimizer $j$ (\textit{mini-j}), and eth=6. The crossbar is divided into three components: (1) Reads FIFO (light yellow), (2) Linear WF Buffer (light blue), and (3) Affine WF Buffer (light green). The numbered circles represent the steps of the entire read-mapping process: (1) indexing -- step \protect\circled{0}, (1) seeding -- step \protect\circled{1}, (2) pre-alignment filtering -- steps \protect\circled{3} and \protect\circled{4}, and (4) read alignment -- step \protect\circled{6}. Steps \protect\circled{2} and \protect\circled{5} involve intra-crossbar data movements, while step \protect\circled{7} entails transmitting the results to the RISC-V. Detailed explanations about this process are given in Sections~\ref{subsec:DART_Indexing} to~\ref{subsec:DART_read_alignment}.
    }
    \label{fig:DART-PIM_WF_within_crossbar}
\end{figure}

Each crossbar (or several crossbars, if needed) handles all read-mapping stages of a specific minimizer from the reference genome.
DART-PIM supports various read lengths, but the best performance is achieved when a single crossbar row accommodates the entire WF matrix calculation (as splitting a read across multiple rows degrades performance).
The row length must, therefore, be at least three times the read size to contain the read (2R bits for R base pairs), the reference (4R bits), and space for intermediate computations/results (minimum $\sim80$ bits).
%Therefore, a crossbar with 1024 columns by 256 rows~\cite{zhang2014selector} supports reads of up to 157 base pairs. Specifically
To support read lengths up to 150 base pairs~\cite{Onur2020review}, each crossbar was allocated 1024 columns by 256 rows~\cite{zhang2014selector}, partitioned into three subsets corresponding to the three online read-mapping stages: seeding, pre-alignment filtering (via linear WF), and read alignment (via affine WF):
\begin{enumerate}
    \item \textit{Reads FIFO} contains 160 rows $\times$ three reads per row (total of $480$ reads associated with the same minimizer).
    \item \textit{Linear WF Buffer} contains 32 rows executing multiple instances of the linear WF concurrently for the target minimizer (one instance per row). Each row accommodates a different segment of the reference.
    \item \textit{Affine WF Buffer} contains 64 rows, where each eight rows facilitate a single affine WF instance (one row for distance calculations and seven rows for optimal alignment recovery). Thus, eight concurrent execution of the affine WF are supported.
\end{enumerate}
Given a target minimizer% (referred to as \textit{mini-j})
, Figure~\ref{fig:DART-PIM_WF_within_crossbar} depicts the crossbar layout and execution flow. The numbering refers to the specific tasks performed in each area.
Due to high variations in minimizer frequency within the reference genome, different minimizers require different compute resources. Infrequent minimizers induce crossbars with low utilization of the linear WF buffer, thus reducing the effective compute per area efficiency. We, therefore, assign the WF calculation of such minimizers to the DP-RISC-V cores offline, saving significant crossbar area.

%This eliminates the need for crossbars to store and compute these minimizers, reducing the required number of crossbars by $36.3\%$ (from 21.1 to 13.4 million crossbars) and increasing linear buffer area efficiency (number of rows containing reference segments, divided by the total number of Linear buffer rows)from $56\%$ to $83\%$. On the other hand, this increases memory usage by $69MB$ because the hash table must be stored in the RISC-V core's DRAM. In this setup, $99.84\%$ of all WF matrices are computed by the memory.

Minimizer frequency within reads also varies significantly. Frequent minimizers fill the Reads FIFO, creating a latency bottleneck. To address this, the Reads FIFO is designed to be significantly larger than the compute buffers. As a result, the number of WF matrices computed for the same minimizer is limited, slightly degrading accuracy (shown to be negligible in  Section~\ref{subsec:results_accuracy}), while dramatically improving execution time.

\subsection{Offline Indexing}
\label{subsec:DART_Indexing}

\begin{figure*}
    \centering
    \includegraphics[trim={0.1cm 0.4cm 0.1cm 0.1cm},height=4cm,clip]{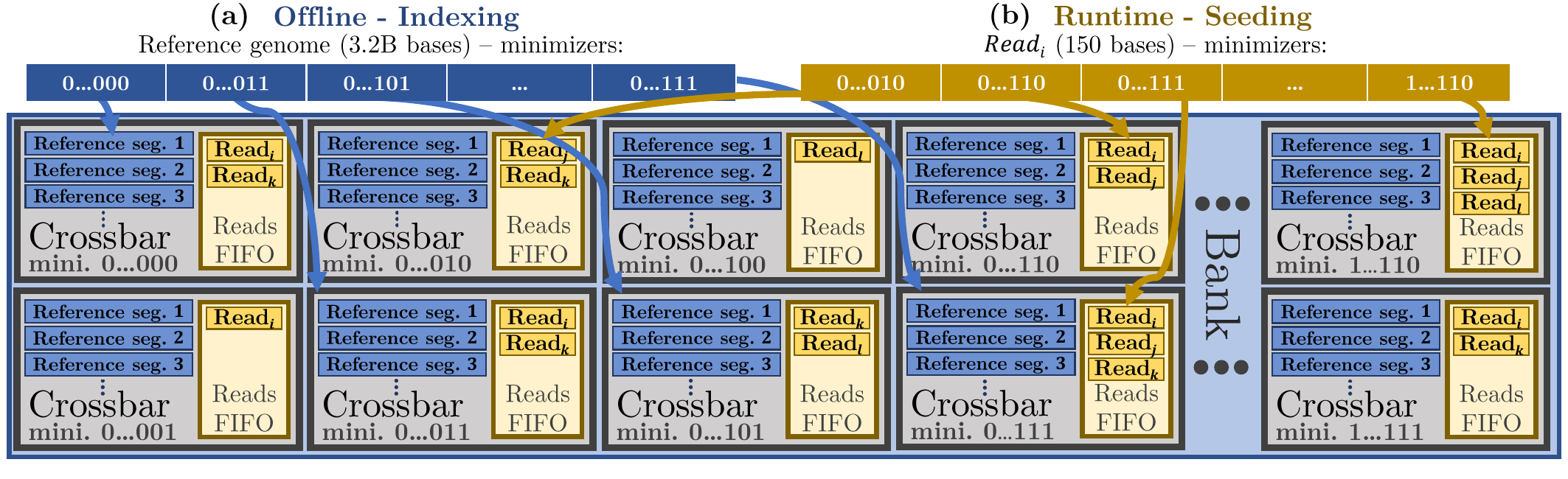}
    \caption{The crossbars' arrangement of (a) the reference genome during the offline indexing stage, and (b) the reads during the runtime seeding stage.}
    \label{fig:DART-PIM_indexing}
\end{figure*}

The offline indexing stage of DART-PIM imports the reference segments themselves (rather than only their addresses) into the linear WF buffers inside the DP-memory crossbars. Each linear WF buffer row is occupied by a single reference segment, representing a single PL, for the minimizer assigned to the relevant crossbar.

The length of the reference segment written to the crossbar is selected to support read alignment regardless of the location of the minimizer within the read. Therefore, the minimum required length is approximately twice the read length, as the minimizer may be at the left/right edge of the read (the precise length is $2(rl+eth)-k$, where \textit{rl} represents the read length).
An illustration of the arrangement of indexing data inside a crossbar is depicted near the \circled{0} marker in Figure~\ref{fig:DART-PIM_WF_within_crossbar}. Figure~\ref{fig:DART-PIM_indexing}a illustrates how indexing is arranged across different crossbars.

Although this approach duplicates the reference segments across crossbars, it is beneficial as it completely eliminates transfers of reference data throughout the computation. Specifically for the human genome, the storage overhead grows about $17\times$, from $800$MB for standard hash-table methods to $13.3$GB for DART-PIM. Our results confirm that this strategy is beneficial in terms of latency and energy, saving around $300$ billion reference segment transfers from the memory.

\subsection{Online Seeding}
\label{subsec:DART_Seeding}

The first online stage of DART-PIM is the seeding of every read to its corresponding reference segments according to the read's set of minimizers. Reads continue to enter the Reads FIFO (\circled{1} in Figure~\ref{fig:DART-PIM_WF_within_crossbar}) until one of the crossbars has filled its Reads FIFO. Figure~\ref{fig:DART-PIM_indexing}b depicts the arrangement of reads across different crossbars according to their minimizers. 

The seeding process commences when the main RISC-V core sends a list of reads and their minimizers to the PIM controllers. Each such controller is aware of all minimizers relevant to any of its descendent controllers in DART-PIM's hierarchy. This way, only relevant reads are propagated throughout DART-PIM to the crossbar controllers, which add them to their Reads FIFO. When one of the Reads FIFO fills up, it signals the PIM controller to stop sending reads; and the same command is sent to the main RISC-V core to stop the read stream and proceed to the pre-alignment filtering stage.

\subsection{Online Pre-Alignment Filtering}
\label{subsec:DART_Filtering}

Pre-alignment filtering in DART-PIM uses the linear WF algorithm. Each read's minimizer is processed by the linear WF to calculate its similarity with all reference segments of its PLs, discarding segments with high distances to reduce the number of PLs per read.

The filtering process begins with the main RISC-V core sending commands to the PIM controller to start WF cell calculations in all DP-memory crossbars. Since the data (read and reference segments) are stored in-memory, the core can send the same instruction to all chip controllers. Each crossbar is then assigned a sequence of MAGIC NOR operations, performing WF matrix cell calculations in its linear WF buffer rows concurrently, as directed by the chip controller—a process we term a \textit{linear WF iteration}.

After each iteration, the processed read is removed from the Reads FIFO. The RISC-V core continues sending reads and repeating the process until all reads are processed and buffers are empty. This stage is computationally intensive and often a bottleneck, making its throughput crucial. Each crossbar sequentially performs the following steps to implement a linear WF iteration, paralleling multiple WF instances:

\begin{enumerate}
    \item First, a read is copied from the Reads FIFO to the crossbar's linear WF buffer (\circled{2} in Figure~\ref{fig:DART-PIM_WF_within_crossbar}). At a given linear WF iteration, a crossbar maps a specific minimizer (within a specific read) to multiple reference segments (representing multiple PLs). Hence, the same read is written to all linear WF buffer rows in parallel.
    Each row performs its computations on a different section of the reference segment, depending on the minimizer's location within the read. This location is given as an address offset to the crossbar controller
    %(depicted in Figure~\ref{fig:crossbar_controll})
    alongside the read itself.
    \item The next stage is the \textit{linear WF matrix calculation} (\circled{3} in Figure.~\ref{fig:DART-PIM_WF_within_crossbar}). The linear WF matrix is computed within the linear WF buffer according to instructions generated by the chip controller for all crossbars in parallel. When this stage concludes, each row contains its linear WF distance.
    \item Consequently, the minimal value is extracted from the distances stored in the different rows of the linear WF buffer (\circled{4} in Figure~\ref{fig:DART-PIM_WF_within_crossbar}). Although computed serially within each crossbar, the minimum extraction is performed in parallel across many crossbars.
    \item Finally, the selected reference segment, corresponding to the minimal WF distance, is copied alongside the read to an empty row in the affine WF buffer (\circled{5} in Figure~\ref{fig:DART-PIM_WF_within_crossbar}).
    Since the read and the reference segment are already aligned at this stage, it is possible to copy only the required sub-segment, rather than the entire reference segment that was twice as long (see Section~\ref{subsec:DART_Indexing}).
\end{enumerate}

\subsection{Online Read Alignment}
\label{subsec:DART_read_alignment}

% \par The read alignment step computes the WF algorithm for hundreds of millions of reads across hundreds of PLs. 

As in the linear WF iteration, an \textit{affine WF iteration} starts within a crossbar only when the Affine WF buffer is full. Once invoked, the affine WF iteration (\circled{6} in Figure~\ref{fig:DART-PIM_WF_within_crossbar}) is calculated in parallel over all crossbar rows as well as across different crossbars. Recall that an affine WF iteration includes both the WF matrix calculation and traceback recovery.

Finally, the results of the affine WF iteration are sent from each crossbar (specifically, each Affine WF buffer row) to the main RISC-V core (\circled{7} in Figure~\ref{fig:DART-PIM_WF_within_crossbar}). The results include the read index, the PL within the reference genome, the affine distance, and the traceback. The RISC-V core maintains in dedicated crossbars a list of reads and their ``best so far'' PL candidate, obtained by selecting the PL with minimal WF distance received from the DP-memory for every read.

%%%%%%%%%%%%%%%%%%%%%%%%%%%%%%%%%%%%%%%%%%%%%%%%%%%%%%%%%%%%%%%%%%%%%%%%
% Evaluations
%%%%%%%%%%%%%%%%%%%%%%%%%%%%%%%%%%%%%%%%%%%%%%%%%%%%%%%%%%%%%%%%%%%%%%%%
\section{DART-PIM Evaluation Methodology}
\label{sec:Evaluation_methodology}

DART-PIM evaluation consists of assessing its accuracy, execution time, energy, and area. We evaluate the results using several simulators and compare them to the following five state-of-the-art platforms: (1) minimap2~\cite{Minimap2}, a CPU-based mapper, executed on an Intel\textregistered\ Xeon\textregistered\ E5-2683 v4 CPU~\cite{CPU_ASIC2_server10} with 32 physical cores operating at $2.10$GHz, with $256$GB DDR4 memory; (2) NVIDIA Parabricks~\cite{parabricks}, a GPU-based mapper, executed on an NVIDIA DGX A100 (with eight A100 Tensor Core GPUs)~\cite{GPU_DGX_A100}; (3) GenASM~\cite{GenASM}, which accelerates computations within the logic layer of a 3D-stacked memory~\cite{HMC,HBM,3DstackedMemory,ahn2015pim}; (4) SeGraM~\cite{SeGraM}, which improves GenASM throughput at the cost of increased energy and area; and (5) GeNVoM~\cite{GeNVoM}, which uses heuristics to improve throughput at the cost of accuracy.

To comprehensively evaluate the DART-PIM architecture, we developed four simulators: (1) a full-system simulator to test the overall behavior and performance; (2) a single-crossbar simulator designed to validate functionality and generate precise metrics for a single WF iteration; (3) a DP-RISC-V simulator for latency evaluation and that includes a main RISC-V module and the remaining RISC-V cores; and (4) synthesized controllers for energy consumption and area evaluation.
The implementations of all simulators, listed below, are accessible in the paper's GitHub repository~\cite{DART-PIM_Github}:

\subsubsection{\textbf{Full-System Simulator}}
An in-house C++ simulator that emulates the entire system operation. It incorporates the full-size PIM memory and executes all (offline and online) stages of read mapping. During the offline phase, the simulator partitions the reference genome into crossbars, as described in Section~\ref{subsec:DART_Indexing}. It then conducts seeding, filtering, and read alignment, as discussed in Sections~\ref{subsec:DART_Seeding},~\ref{subsec:DART_Filtering}, and~\ref{subsec:DART_read_alignment}. The simulator also provides the exact number of linear and affine WF instances and iterations performed by DART-PIM throughout the process. Additionally, it quantifies the total utilized memory capacity and the resulting read-mapping accuracy.

\subsubsection{\textbf{Single-Crossbar Simulator}} 
An in-house MATLAB \cite{Matlab} simulator that models a single crossbar operation cycle by cycle. This simulator verifies the functionality of algorithms implemented using MAGIC NOR operations within DART-PIM. It provides an exact execution time by computing the number of cycles required for executing linear and affine WF, including pre- and post-processing operations. Additionally, it quantifies the energy consumption by counting the number of write/MAGIC NOR switches and the number of read bits.

\subsubsection{\textbf{RISC-V Simulator}} 
A GEM5~\cite{gem5} based simulator, developed to simulate the DART-PIM RISC-V cores (\textit{DP-RISC-V}). A designated module emulates the main RISC-V core, while the other RISC-V cores, used for computing tasks, are modeled by other modules, each with a cache memory. This simulator estimates the execution time of each RISC-V core within the DP-RISC-V, emulating the WF computation distribution between the DP-RISC-V and DP-memory, as well as the communication between them.

\subsubsection{\textbf{PIM/Chip/Bank/Crossbar Controllers}} 
We synthesized the controllers included in DART-PIM, as depicted in Figure~\ref{fig:DART_arch}, to estimate their area and power consumption. The controllers were designed and implemented in SystemVerilog and synthesized using Synopsys Design Compiler~\cite{DesignCompiler}, targeting TSMC 28nm CMOS technology.

Our evaluation datasets comprise the full Homo Sapiens (human) reference genome (GRCh38\_latest\_genomic.fna~\cite{NCBI}) along with short-read datasets generated by Illumina HiSeq X~\cite{Illumina_HiSeqX} (HG002.hiseqx.pcr-free.30x.R[1-2].fastq~\cite{NCBI}). These datasets comprise $389M$ reads, each 150 base pairs long.

\section{Experimental Results}
\label{sec:Evaluation}

\begin{table}[]
    \centering
    \caption{Summary of DART-PIM used parameters.}
    \label{tab:dart-pim-params}
    \scriptsize
    \begin{tabular}{|c|c|c|}
        \hline
        \textbf{Parameter} & \textbf{Context} & \textbf{Value} \\
        \hline
        Read length ($rl$) & Read mapping & $150$ \\ 
        \hline
        Minimizer length ($k$) & Read mapping & $12$ \\ 
        \hline
        Minimizer window length ($W$) & Read mapping & $30$ \\ 
        \hline
        Linear (affine) error threshold ($eth$) & Read mapping & $6$ ($31$) \\ 
        \hline
        $w_{sub}$=$w_{ins}$=$w_{del}$=$w_{op}$=$w_{ex}$ & Wagner-Fischer & 1 \\ 
        \hline
        Reads FIFO $\#$ rows & DART-PIM & $160$ \\
        \hline
        Linear Buffer $\#$ rows & DART-PIM & $32$ \\
        \hline
        Affine Buffer $\#$ rows & DART-PIM & $64$ \\
        \hline
        Low threshold (lowTh) & DART-PIM & $3$ \\
        \hline
        Maximum $\#$ allowed read (max reads) & DART-PIM & $25$k \\
        \hline
        Low cell resistance ($\text{R}_\text{on}$) & RRAM & $50\text{k}\Omega$ \\
        \hline
        High cell resistance ($\text{R}_\text{off}$) & RRAM & $5\text{M}\Omega$ \\
        \hline
    \end{tabular}
\end{table}

The evaluation of DART-PIM is based on extensive simulations performed using the simulators outlined in Section~\ref{sec:Evaluation_methodology} and parameters listed in Table~\ref{tab:dart-pim-params}.
General parameters (i.e., read-mapping and Wagner-Fischer parameters) were set according to existing literature and remain constant across all simulations.
% Future advancements may expand the physical availability of columns beyond 1024, enabling the mapping of longer reads. 
% %As the size of each crossbar increases, the read length grows linearly with the size of the row. 
% For instance, with row lengths of 2048 (4096), reads of up to 300 (600) base pairs can fit within a single memory row. Longer reads can be divided into multiple rows, with execution time scaling with the number of rows.
Parameters used specifically in DART-PIM were selected using an exhaustive search to optimize system latency and energy consumption. The linear (affine) buffer size was set to 32 (64) rows, resulting in a Reads FIFO of $256-32-64 = 160$ rows (recall the $256 \times 1024$ crossbar size). 
% A fine-tuning process was conducted to optimize DART-PIM other parameters, focusing on maximizing area utilization and minimizing execution time and energy consumption. Details of this process can be found in Appendix B~\cite{DART-PIM_Github}. The chosen parameters are as follows: linear and affine buffer sizes are set to 32 and 64 rows, respectively, resulting in a Reads FIFO with $256-32-64 = 160$ rows. 

Another important parameter is the low threshold $\textit{lowTh}=3$. The low threshold balances between DP-memory capacity and RISC-V workload by determining whether affine WF computations are performed by DP-memory (minimizer frequency above \textit{lowTh}) or DP-RISC-V (minimizer frequency below or equal \textit{lowTh}).
To avoid highly non-uniform assignment of reads per crossbar, we also bound the number of read handled by any single crossbar to \textit{maxReads} (=$12.5$k/$20$k/$50$k). %This value was found to yield a reasonable trade-off between execution time and accuracy impact.
% To avoid highly non-uniform assignment of reads per crossbar, we bound the number of read handled by any single crossbar to be \textit{maxReads}=50K, thereby controlling the maximal execution time (at the potential cost of accuracy reduction).
% The description of the parameter search is included in the paper's GitHub repository~\cite{DART-PIM_Github}.

This section details the simulated results of DART-PIM's accuracy, execution time, energy consumption, and area, and compares them to minimap2~\cite{Minimap2}, NVIDIA Parabricks~\cite{parabricks}, GenASM~\cite{GenASM}, SeGraM~\cite{SeGraM} and GeNVoM~\cite{GeNVoM}.

\subsection{DART-PIM Accuracy}
\label{subsec:results_accuracy}

%Before evaluating all DART-PIM characteristics, we verify that results accuracy is not compromised. To evaluate the accuracy of DART-PIM, 
To evaluate DART-PIM's accuracy, we compare it to \textit{BWA-MEM}~\cite{BWA-MEM} that serves as the ground-truth. Specifically, we compare the locations identified by DART-PIM to those selected by BWA-MEM.
Unlike GenASM and SeGraM which use score similarity to approximate accuracy, our accuracy metric is defined as the fraction of DART-PIM mappings that precisely match the corresponding BWA-MEM mapped locations, and as such, it is more meaningful.
%Despite simplifications and heuristics employed by DART-PIM, such as sub-optimal pre-alignment filtering and discard of some frequent read minimizers, its accuracy reaches 
%. Our evaluation demonstrates accuracy comparable to state-of-the-art computationally-intensive approaches 
DART-PIM has an accuracy of $99.7\%$ for maxReads of $12.5$k, and $99.8\%$ for maxReads of $25$k and $50$k, while Parabricks and minimap2 each have an accuracy of $99.9\%$~\cite{parabricks,Minimap2}. GenASM and SeGraM both achieve $96.6\%$ accuracy~\cite{GenASM,SeGraM}, and GenVoM has an accuracy of $91.2\%$ due to its use of heuristics~\cite{GeNVoM}.

\begin{figure}
    \centering
    \includegraphics[width=.8\linewidth]{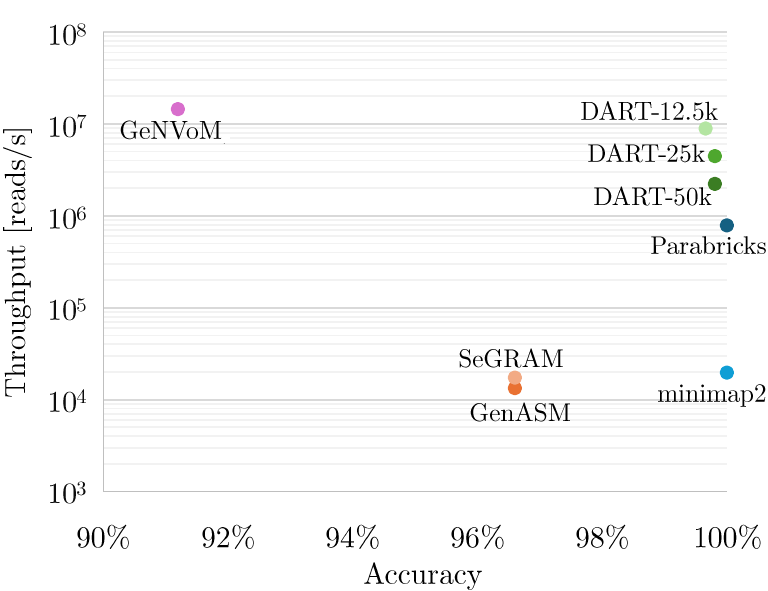}
    \caption{A comparison of DART-PIM with prior read-mapping accelerators in terms of throughput and accuracy.}
    \label{fig:AccuracyVsThroughput}
\end{figure}

Figure~\ref{fig:AccuracyVsThroughput} demonstrates the trade-off between an accelerator’s throughput and the achieved mapping accuracy. DART-PIM stands out by optimizing this trade-off, attaining throughput slightly below the state-of-the-art (GeNVoM) while maintaining state-of-the-art accuracy (Parabricks/minimap2). DART-PIM is positioned similarly when considering the trade-off between accuracy and area/energy efficiencies.

\subsection{Single Crossbar Evaluation}
\label{subsec:results_single_crossbar}

\begin{table}[]
\scriptsize
\centering
\caption{Cycle and switch counts for each memory operation from the single-crossbar simulator during a single WF calculation. %It includes the computation of the minimum value among all simultaneously calculated WFs within the crossbar.
%There are no reads during the computations, so the table does not pertain to them.
}
\label{tab:singleWF}
\setlength{\tabcolsep}{4pt}
\begin{tabular}{|c|c|c|c|c||c|}
\hline
& \textbf{} & \textbf{MAGIC NOR} & \textbf{Writes} & \textbf{Reads} & \textbf{Total}\\ \hline
\multirow{2}{*}{\textbf{\begin{tabular}[c]{@{}c@{}}Linear \\ WF\end{tabular}}} & \textbf{\# Cycles} & 254,585 &  4035 & 0 & 258,620 \\ \cline{2-6} 
& \textbf{\# Switches}  &  254,384 &  255,499  & 0 & 509,883      \\ \hline
\multirow{2}{*}{\textbf{\begin{tabular}[c]{@{}c@{}}Affine\\ WF\end{tabular}}} & \textbf{\# Cycles} &  1,288,281 &  20,418 & 0 & 1,308,699 \\ \cline{2-6}  & \textbf{\# Switches}  &  1,271,921  &  1,277,495 & 0 & 2,549,416
\\ \hline
\end{tabular}

\end{table}

\begin{table}[]
    \centering
    \caption{Summary of conservatively scaled MAGIC NOR and write switching energy and cycle time from~\cite{RACER2021}.}
    \label{tab:Energy&Times}
    \scriptsize
    \begin{tabular}{|c|c|}
        \hline
        \textbf{Operations} & \textbf{Energy/Time} \\ 
        \hline
        MAGIC/write conservatively scaled cycle & 2ns \\ 
        \hline
        MAGIC conservatively scaled energy & 90fJ/bit \\ 
        \hline
        Write conservatively scaled energy & 90fJ/bit  \\ 
        \hline  
    \end{tabular}
\end{table}

The fundamental building block of our performance evaluation is a cycle-accurate simulation of executing WF instances on a single crossbar. The Single-Crossbar simulator counts the overall number of MAGIC NOR switches, number of read bits and write switches, and the number of cycles per operation, encompassing both linear and affine algorithms.

The number of cycles and switches utilized by a single WF instance is outlined in Table~\ref{tab:singleWF}. Thanks to inter- and intra-crossbar parallelism, multiple WF instances are performed at the same number of cycles. The energy consumption of nucleus operations (write / MAGIC NOR switches), required for calculating the total energy per WF instance, appear in Table~\ref{tab:Energy&Times}. Given these values, the number of cycles required for the linear (affine) WF is $258,620$ ($1,308,699$), while the overall energy consumed by the linear (affine) WF amounts to $509,883 \times 90fJ = 45.9nJ$ ($2,549,416 \times 90fJ = 229nJ$). 

The simulated number of cycles ($254,585$) fits our theoretical estimates: the total MAGIC NOR cycles for a single linear WF computation is the product of $1950$ WF cells ($2eth+1 = 13$ cells per row times $150$ rows) and $130$ cycles per cell, which amounts to $253,500$ cycles. The remaining $1085$ cycles are dedicated to the first matrix row/column initializations, and to step \circled{4} in Figure~\ref{fig:DART-PIM_WF_within_crossbar}.

%add same about affine?  - no space...
%throughput of a single crossbar = #instances/latency = 16/258,620 = 

\begin{figure*}
    \centering
    \includegraphics[width=1\linewidth]{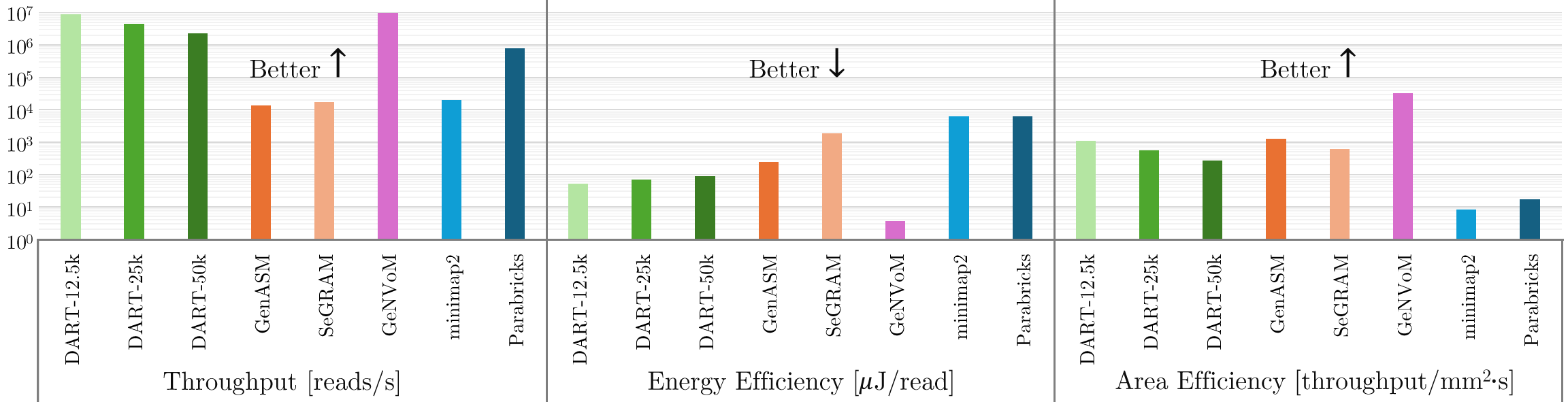}
    \caption{Comparison of throughput (reads per second), energy efficiency (read per joule), and area efficiency (throughput per squared millimeter) of DART-PIM with different values of maxReads ($12.5$k, $25$K, $50$K) and state-of-the-art approaches.}
    \label{fig:StateOfTheArtComparison}
\end{figure*}

\begin{figure*}[h]
\centering
  \begin{subfigure}{.32\textwidth}
  \centering\captionsetup{width=.95\linewidth}
    \includegraphics[width=.95\linewidth]{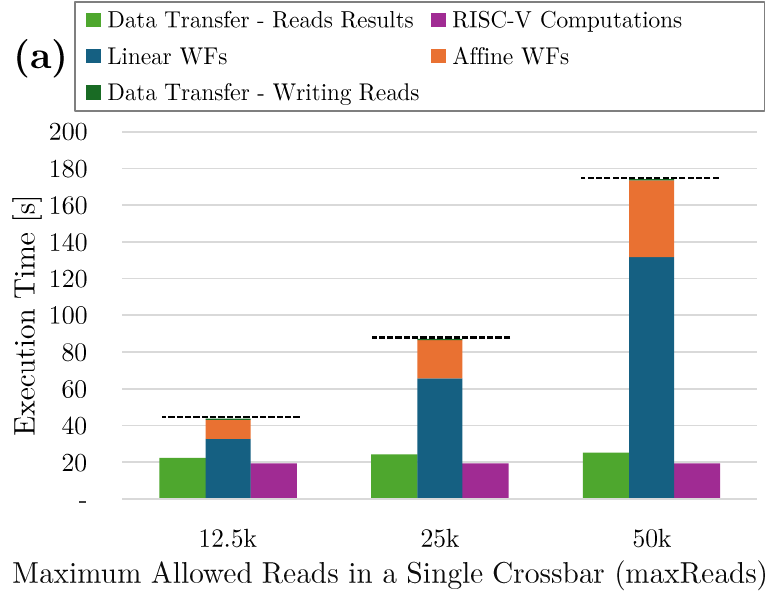}
    %\caption{Execution time across various maxReads settings, alongside time of the data transfer and DP-RISC-V computations.}
    %\label{fig:execution_time}
  \end{subfigure}%
  \vspace{15pt}
  \begin{subfigure}{.32\textwidth}
  \centering\captionsetup{width=.95\linewidth}
    \includegraphics[width=.95\linewidth]{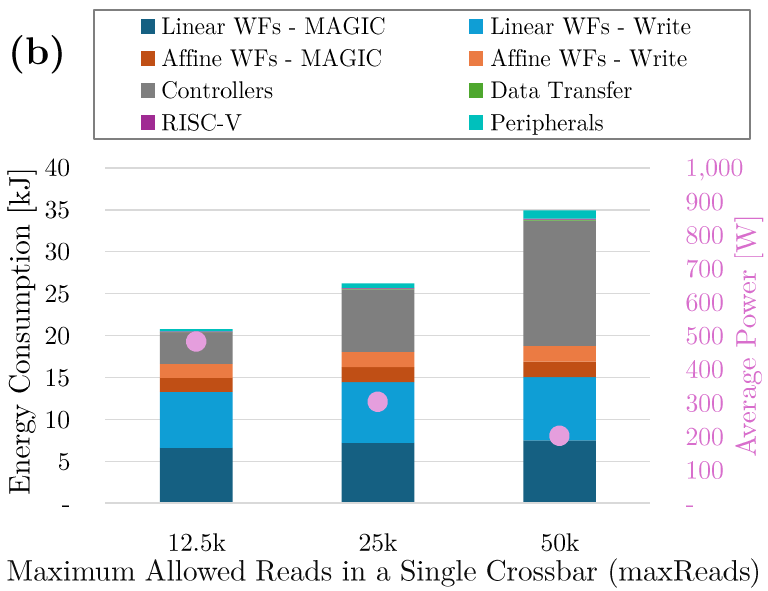}
    %\caption{Energy consumption and average power across different maxReads configurations.\\}
    %\label{fig:energy_consumption}
  \end{subfigure}%
  \vspace{15pt}
   \begin{subfigure}{.32\textwidth}
   \centering\captionsetup{width=.95\linewidth}
    \includegraphics[width=.95\linewidth]{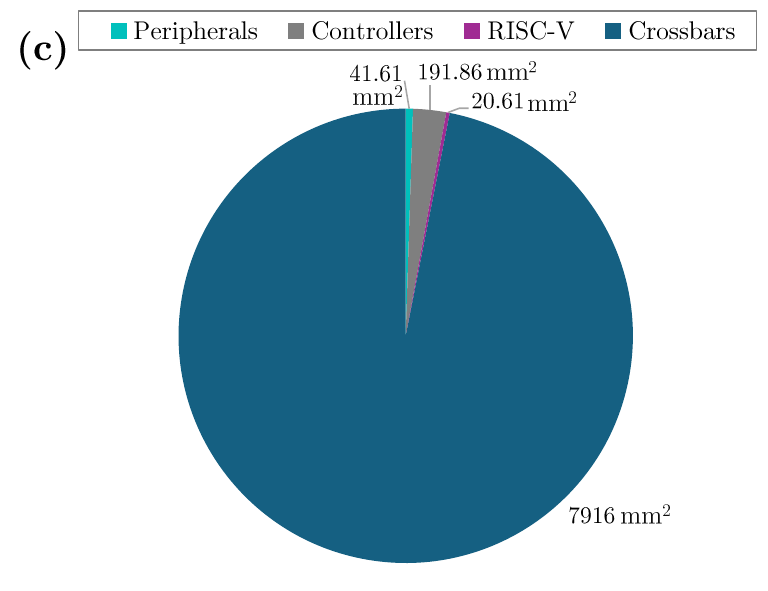}
    %\caption{Area of crossbars, periphery, controllers and RISC-V.\\}
    %\label{fig:area}
  \end{subfigure}
\vspace{-30pt}
\caption{Performance of DART-PIM for $389$ million reads: (a) execution time across various maxReads settings, alongside time of the data transfer and DP-RISC-V computations, (b) energy consumption and average power across different maxReads configurations, and (c) area of crossbars, periphery, controllers, and RISC-V.}
\label{fig:Graphs}
\end{figure*}

\subsection{DART-PIM Execution Time and Throughput}
\label{subsec:results_time}

Figure~\ref{fig:StateOfTheArtComparison} (left subplot) compares the throughput of DART-PIM, minimap2~\cite{Minimap2}, NVIDIA Parabricks~\cite{parabricks}, GenASM~\cite{GenASM}, SeGraM~\cite{SeGraM}, and GeNVoM~\cite{GeNVoM}, measured by mapped reads per second. 
For our selected dataset ($389M$ reads), DART-PIM execution time ranges between $43.8$s (maxReads~=~$12.5$k) to $174$s (maxReads~=~$50$k). In comparison, minimap2 takes $5.5$ hours ($19,785$s) on a Xeon E5-2683 v4, while Parabricks completes in $8.3$ minutes ($495$s) on DGX A100. GenASM requires $8.1$ hours ($29,154$s), based on the reported $200$k reads per $30$s. This time is scaled from the original GenASM results with a read length of 250 base pairs to 150 base pairs, according to the throughput ratio of those read lengths. SeGraM shows a $1.3\times$ throughput improvement over GenASM, resulting in an execution time of $6.23$ hours ($22,426$s). According to GenVoM's reported throughput, its execution time for our dataset is $39.2$s (scaled from reads of 100 to 150 base pairs).
When using maxRead~=~$25$k, DART-PIM's execution time compared to minimap2, Parabricks, GenASM, and SeGraM are $227\times$, $5.7\times$, $334\times$ and $257\times$ better, respectively.

Figure~\ref{fig:Graphs}a shows the breakdown of DART-PIM's execution time. The execution time is determined by the maximum of three latency factors: (1) Reading the results from the DP-memory (light green), (2) overall latency of DP-RISC-V computations (purple), and (3) the time required to write the reads to DP-memory and compute within the DP-memory (aggregate of remaining colors). To ensure that DART-PIM is primarily constrained by the computational capabilities of PIM rather than being limited by data transfers or RISC-V processing, the latter should be the most time-consuming.

The DP-memory execution time is determined as

{\footnotesize
\begin{equation}
    T_{\textnormal{DPmemory}}=(K_L \cdot N_L + K_A \cdot N_A) \cdot T_{clk},
\end{equation}}
where $T_{clk}$ is the clock cycle time (see Table~\ref{tab:Energy&Times}), $K_L$ ($K_A$) is the number of linear (affine) iterations (determined by the full-system simulator), and $ N_L $ ($ N_A $) is the number of cycles per linear (affine) WF iteration (see Table~\ref{tab:singleWF}).

The maximal number of reads per crossbar trades off accuracy and execution time: A higher maxReads enhances accuracy, but linearly increases total execution time. As the number of linear/affine WF instances grows sub-linearly with maxReads, the throughput (reads per second) decreases.

We chose to add $128$ RISC-V cores (4 per chip) to guarantee that DP-RISC-V do not pose a bottleneck. Assuming all cores are working in parallel, we evaluated their execution time on $0.16\%$ of all affine WF instances to be $19.4$s. This underlines the extreme efficiency of DP-memory, as it computes $99.84\%$ of the instances in only four times this latency.

\subsection{DART-PIM Energy Efficiency}
\label{subsec:results_energy}

Figure~\ref{fig:StateOfTheArtComparison}~(middle subplot) compares the energy efficiency~\cite{DarkMemory} of DART-PIM, minimap2, NVIDIA Parabricks, GenASM, SeGraM, and GeNVoM, measured as the consumed energy per mapped read. For our selected dataset ($389$M reads), minimap2 and Parabricks both consume $2.4$MJ of energy, based on average power consumption of $120$W and $4850$W, respectively.
The energy consumption for GenASM, SeGraM, and GeNVoM is $94.2$kJ, $543$kJ, and $1.4$kJ, respectively, based on average power consumption of $3.23$W, $24.2$W ($7.5\times$ GenASM's power), and $35.3$W.  
DART-PIM consumes between $20.8$kJ (for maxRead~=~$12.5$k) and $34.9$kJ (for maxRead~=~$50$k), with average power consumption ranging from $201$W to $482$W). For maxRead~=~$25$k, DART-PIM achieves a respective improvement of $90.6$x over minimap2 and Parabricks, $3.6$x over GenASM, and $20.7$x over SeGraM.

% The energy consumed by DART-PIM significantly outperforms both NVIDIA Parabricks implementation and state-of-the-art GenASM approaches. First, DART-PIM energy consumption evaluation is presented, and then a comparison to state-of-the-art is shown.

Figure~\ref{fig:Graphs}b shows a breakdown of DART-PIM's energy consumption to three primary components: DP-memory (controller, peripheral circuits, and crossbars), DP-RISC-V (cores and caches), and the mutual data transfer. 
Given an energy consumption of $\mathcal{E}_{MAGIC}$ and $\mathcal{E}_{WRITE}$ for MAGIC and WRITE operations, respectively, the crossbars energy is determined by

{\footnotesize
\begin{equation}
\begin{aligned}
    & \mathcal{E}_{\textnormal{crossbars}}=
(\mathcal{E}_{\textnormal{MAGIC}} \cdot S^{\textnormal{MAGIC}}_L + \mathcal{E}_{\textnormal{WRITE}} \cdot S^{\textnormal{WRITE}}_L) \cdot J_L
    \\ & + 
    (\mathcal{E}_{\textnormal{MAGIC}} \cdot S^{\textnormal{MAGIC}}_A + \mathcal{E}_{\textnormal{WRITE}} \cdot S^{\textnormal{WRITE}}_A) \cdot J_A,
\end{aligned}
\end{equation}}
% where $S^{MAGIC}_L$ ($S^{MAGIC}_A$) is the number of linear (affine) MAGIC switches, $S^{WRITE}_L$ ($S^{WRITE}_A$) is the number of linear (affine) write switches, and $J_L$ ($J_A$) is the number of linear (affine) instances. The parameters' values appear in Tables~\ref{tab:Energy&Times},~\ref{tab:singleWF}.
where $S^{MAGIC}_L$, $S^{WRITE}_L$ and $J_L$ are the number MAGIC switches, write switches, and instances for linear WF, respectively ($S^{MAGIC}_A$, $S^{WRITE}_A$ and $J_A$ are the corresponding parameters for affine WF). Their values appear in Tables~\ref{tab:Energy&Times},~\ref{tab:singleWF}.

\begin{table}[]
    \centering
    \caption{Time, energy, and area of data transfer, RISC-V cores, peripherals, and controllers (of a single unit). [Cs]/[RVs] indicate data generated by the controller/RISC-V simulator.}
    %\caption{Time, energy, and area of data transfer, RISC-V cores, peripherals, and controllers (of a single unit). [Cs] indicates data generated by the controller simulator and [RVs] denotes data from the RISC-V simulator.}
    \label{tab:Energy&TimesPeripherals}
    \tiny
    \begin{tabular}{|c||c|c|c|c|}
    
        \hline
        \textbf{Operations} & \textbf{Time} & \textbf{Energy/Power}  & \textbf{Area} & \textbf{\# Units}\\ 
        \hline
        Write (DP-RISC-V to DP-memory) & 32GB/s~\cite{RACER2021} & 11.7pJ/bit~\cite{Nishil_CONCEPT} & - & - \\ %(energy - memory array + periphery) 
        \hline
        Read (DP-memory to DP-RISC-V) & 32GB/s~\cite{RACER2021} & 5.64pJ/bit~\cite{Nishil_CONCEPT} & - & -\\  %(energy - memory array + periphery) 
        \hline
        RISC-V (single affine SW instance)  & 88$\mu$s  [RVs] &  40mW~\cite{RISC-V_andrey} &  0.11mm\textsuperscript{2}
        \cite{RISC-V_andrey} & 128 \\ 
        \hline
        RISC-V cache  & - &  8mW~\cite{RISC-V_cache} &  0.05mm\textsuperscript{2}
        \cite{RISC-V_cache} & 128 \\ 
        \hline
        Crossbar controller [Cs] & - &  9.43$\mu$W & 21$\mu$m\textsuperscript{2} &  8M 
        \\ 
        \hline
        Bank controller [Cs] & - &  0.42mW & 939$\mu$m\textsuperscript{2} &  16k 
        \\ 
        \hline
        Chip controller [Cs] & - & 9.4mW & 20,091$\mu$m\textsuperscript{2} & 16\\ 
        \hline
        PIM controller [Cs] & - &  0.5mW & 938$\mu$m\textsuperscript{2} & 1\\ 
        \hline
        Decode and Drive Unit~\cite{RACER2021} & - &  129.1$\mu$W & 277$\mu$m\textsuperscript{2} & 16k\\ 
        \hline
        R/W Circuit~\cite{RACER2021} & - &  10pW & 0.06$\mu$m\textsuperscript{2} & 8M \\ 
        \hline
        Selector Passgates~\cite{RACER2021} & - &  20pW & 0.001$\mu$m\textsuperscript{2} & 1024$\times$8M \\ 
        \hline
        Driver Passgate~\cite{RACER2021} & - &  20pW & 0.001$\mu$m\textsuperscript{2} & 256$\times$8M \\ 
        \hline
    
    \end{tabular}

\end{table}

The power analysis of the RISC-V cores, detailed in Table~\ref{tab:Energy&TimesPeripherals}, was conducted using AndesCore AX25~\cite{RISC-V_andrey}. All RISC-V cores with their caches consume $6.1$W of power, as shown in Figure~\ref{fig:Graphs}b (purple). Based on the controller configuration (Table~\ref{tab:configurations}) and corresponding controller simulation results (Table~\ref{tab:Energy&TimesPeripherals}), the aggregated average power consumption is $86$W, yielding the energy consumption in Figure~\ref{fig:Graphs}b (grey).

The remaining energy consumption corresponds to memory peripherals and data transfers. The peripheral circuits were evaluated by circuit synthesis analysis conducted by RACER~\cite{RACER2021} (scaled to $28$nm) and amounted to $5.7$W (see details in Table~\ref{tab:Energy&TimesPeripherals}). The data transfer energy comprises the writeout of reads to DP-memory ($1.1$J) and readout of computation results from the DP-memory ($75.4$J), as evaluated by CONCEPT~\cite{Nishil_CONCEPT}. Figure~\ref{fig:Graphs}b presents the estimates for memory peripherals (turquoise) and data transfers (green).

Increasing maxReads results in linear growth of execution time and in the controllers' and peripheral devices' energy consumption, while the power consumption remains relatively stable.
The energy consumption for DP-memory computations, however, rises only from $16.6$kJ at maxReads~=~12.5k to $18.8$kJ at maxReads~=~50k. This increase corresponds to the marginal growth in WF instances computed by DP-memory.

\subsection{DART-PIM Area Efficiency}
\label{subsec:results_area}

Figure~\ref{fig:StateOfTheArtComparison} (right subplot) compares the area efficiency~\cite{DarkMemory} of DART-PIM, minimap2, NVIDIA Parabricks, GenASM, SeGraM, and GeNVoM, evaluated as the read-mapping throughput per chip area. DART-PIM's area efficiency spans between $1086 \frac{\textnormal{reads}}{\textnormal{mm}^2\cdot s}$ for maxReads=$12.5$k and $273 \frac{\textnormal{reads}}{\textnormal{mm}^2\cdot s}$ for maxReads=$50$K. This positions DART-PIM close to GenASM's and SeGraM's $1247 \frac{\textnormal{reads}}{\textnormal{mm}^2\cdot s}$ and $623 \frac{\textnormal{reads}}{\textnormal{mm}^2\cdot s}$, respectively, and significantly better than minimap2's and NVIDIA Parabricks' respective $8.3 \frac{\textnormal{reads}}{\textnormal{mm}^2\cdot s}$ and $16.9 \frac{\textnormal{reads}}{\textnormal{mm}^2\cdot s}$. 

We used the reported areas of $10.7 \textnormal{mm}^2$ for GenASM, $27.8 \textnormal{mm}^2$ ($2.6\times$ GenASM’s area) for SeGraM~\cite{GenASM,SeGraM}, $298 \textnormal{mm}^2$ for GeNVoM~\cite{GeNVoM}, and $2,362 \textnormal{mm}^2$ for the CPU used by minimap2~\cite{CPU_ASIC2_server10}. The area of Parabricks was calculated considering eight A100 GPUs, each combined with six HBMs stacks of nine 8Gb (eight units and a 8Gb buffer die), each with an area of 92mm\textsuperscript{2}~\cite{HbmUnitSize,HbmSize}. 
The total area of A100 GPUs and their HBMs is 
$8 \cdot ( 826 \textnormal{mm}^2 + 6 \cdot 9 \cdot 92 \textnormal{mm}^2 ) = 46,352\textnormal{mm}^2$.

Aggregating its four components, DART-PIM's total area amounts to $8170 \textnormal{mm}^2$: (1) crossbar arrays, (2) controllers, (3) memory peripherals, and (4) RISC-V units. 
Figure~\ref{fig:Graphs}c shows a breakdown of DART-PIM's area, illustrating that crossbars occupy $96.9\%$ of the area. Since all crossbars performs similar tasks, we were able to substantially simplify the CMOS-based crossbars controllers and save significant chip area.

The areas of all DART-PIM's components are listed in Table~\ref{tab:Energy&TimesPeripherals} for CMOS $28$nm technology. Specifically, the area of $128$ DP-RISC-V cores and caches amounts to $14.2\textnormal{mm}^2$ (based on AndesCore AX25~\cite{RISC-V_andrey}) and $6.4\textnormal{mm}^2$, respectively; the area of the controllers sums up to $191.9\textnormal{mm}^2$ (determined by synthesis results); and the peripherals occupy $53.6\textnormal{mm}^2$ (based on $15$nm synthesis from RACER~\cite{RACER2021} using the Synopsis Design Compiler, scaled to the target $28$nm technology~\cite{scaling}).
The remaining area of DART-PIM comprises $8M$ crossbars, as extracted from the full-system simulator. Based on~\cite{hsieh2013ultra}, each crossbar cell has a feature size ($F$) of $30$nm, resulting in a memory cell area of $4F^{2} = 3600\textnormal{nm}^2$. With a crossbar size of $256\times1024$, the area per crossbar is $944 \mu \textnormal{m}^{2}$ (and $7916\textnormal{mm}^2$ for all crossbars).
Note that the total memory capacity of DART-PIM is  $8M \times 1024 \times 256$ resulting in $256GB$. This confirms the practicality of DART-PIM as it aligns with common memory DIMMs, such as the Intel Optane persistent memory~\cite{Intel_optane}.

\section{Related Work}
\label{sec:rel_work}

To the best of our knowledge, DART-PIM represents the first attempt to improve the entire read mapping by eliminating data movement between the different read mapping stages. By consolidating all stages of read mapping within the same physical memory, PIM effectively eliminates the need for external data transfer. Recent research efforts have primarily focused on accelerating a part of the read mapping process.

Various attempts focused on solely accelerating the indexing and/or filtering stages. Indexing was accelerated both over classical architectures, e.g. GPU~\cite{chacon2014indexing}, FPGA~\cite{Yang2023indexing} and ASIC (Niubility~\cite{Niubility} and EXMA~\cite{EXMA}), and using near-memory processing~\cite{herruzo2021indexing} or PIM (FindeR~\cite{FindeR},~AligneR~\cite{Aligner}).
Filtering acceleration was also addressed in previous work, including FastHASH~\cite{fastHASH} for standard CPUs, GRIM Filter~\cite{GRIMfilter} for 3D-stacked memory, and embedded dynamic memory based RASSA~\cite{kaplan2018rassa}. Additionally, FPGA-based solutions (e.g., MAGNET~\cite{MAGNET}, GateKeeper~\cite{GateKeeper} and Shouji~\cite{shouji}), alongside PIM and in-storage frameworks, (e.g., FiltPIM~\cite{FiltPIM} and GenStore~\cite{GenStore}), leverage parallelism to expedite the filtering process.

Different approaches utilize specialized software or hardware accelerators tailored to the demanding read alignment process. Noteworthy software solutions include Minimap2~\cite{Minimap2} and BWA-MEM~\cite{BWA-MEM}. On the hardware side, Parasail~\cite{Parasail} leverages SIMD-capable CPUs, while GPU-RMAP~\cite{GPU-RMAP} and GSWABE~\cite{GSWABE} utilize GPUs. FPGAs were employed by FPGASW~\cite{FPGASW} and ASAP~\cite{ASAP}, while ASIC-based accelerators have also been proposed~(Darwin~\cite{darwin2018}). 
Moreover, GenDP~\cite{GenDP} suggests a DP accelerator with programmable compute units.
Recently, PIM-based solutions such as~\cite{kaplan2017resistive}, RADAR~\cite{RADAR},
RAPID~\cite{RAPID_Gupta}, RAPIDx~\cite{RAPIDx}, and BioSEAL~\cite{Bioseal} offer promising alternatives for more efficient read alignment.

The remainder of relevant prior work includes acceleration of all read-mapping stages. GenAx~\cite{GenAx} and its improved version GenCache~\cite{GenCache} accelerate each stage individually, but do not consider data transfer in between stages in their performance evaluation. GenPIP~\cite{GenPIP}, on the other hand, integrated read mapping with a preliminary process (called basecalling), and evaluated their acceleration altogether. For a coherent evaluation, we compare DART-PIM only to architectures that accelerate and evaluate the end-to-end read-mapping process: GenASM~\cite{GenASM}, SeGraM~\cite{SeGraM}, GeNVoM~\cite{GeNVoM}.

While GenASM~\cite{GenASM} markedly enhances both filtering and read alignment  within the logic layer of a
3D-stacked memory, its reliance on data transfers between different stages restricts overall performance improvement. 
SeGraM~\cite{SeGraM} improved GenASM by introducing an algorithm/hardware co-design for improving the seeding and read alignment stages, resulting in a throughput enhancement at the cost of energy and area.
Finally, GeNVoM~\cite{GeNVoM} employs an associative memristive memory to implement a heuristic search-based procedure, instead of sequence alignment methods (used in this paper). The main problem of this approach is a severe accuracy degradation.
In contrast to above solutions, DART-PIM mitigates the data transfer bottleneck and enables very high parallelism by implementing all stages of read mapping in-memory without sacrificing mapping accuracy.

%%%%%%%%%%%%%%%%%%%%%%%%%%%%%%%%%%%%%%%%%%%%%%%%%%%%%%%%%%%%%%%%%%%%%%%%
% Conclusions
%%%%%%%%%%%%%%%%%%%%%%%%%%%%%%%%%%%%%%%%%%%%%%%%%%%%%%%%%%%%%%%%%%%%%%%%
\section{Conclusion}
\label{sec:conclusion}

We introduce DART-PIM, a novel framework for accelerating the entire DNA read-mapping process using digital PIM. DART-PIM integrates all stages of read mapping within the same physical crossbars, eliminating the performance and energy costs associated with data transfers. Our evaluation shows that memristive-based DART-PIM achieves significant speedup and improved energy efficiency compared to state-of-the-art existing accelerator and GPU implementations.

The DART-PIM concept and architecture are not limited to RRAM or memristive logic and could be adapted to leverage other PIM techniques, including those based on DRAM, SRAM, and different emerging PIM technologies. 

% We envision DART-PIM as a catalyst for inspiring future research into leveraging PIM for read mapping and other bioinformatics workloads, intending to optimize data organization to eliminate the need for data transfers altogether.

%%%%%%% -- PAPER CONTENT ENDS -- %%%%%%%%

%%%%%%%%% -- BIB STYLE AND FILE -- %%%%%%%%
\bibliographystyle{IEEEtranS}
\bibliography{refs}
%%%%%%%%%%%%%%%%%%%%%%%%%%%%%%%%%%%%

\end{document}